\documentstyle[11pt,aas2pp4,epsfig]{article}

\newcommand{\zsol}{$Z_{\odot}$}
\newcommand{\NH}{$N_{\rm H}$}
\newcommand{\dof}{d.o.f.}
\newcommand{\znf}{$Z_{\rm NF}$}
\newcommand{\zomss}{$Z_{\rm OMSS}$}
\newcommand{\zhcn}{$Z_{\rm NACN}$}

\tighten

\lefthead{K. Weaver et al.}
\righthead{X-ray properties of NGC\,253 and M\,82}

\begin{document}
 
\title{An X-ray Mini-survey of Nearby Edge-on Starburst Galaxies \\ II.
The Question of Metal Abundance}

\author{Kimberly A. Weaver}
\affil{Code 662 NASA/Goddard Space Flight Center, Greenbelt MD 20771;
Johns Hopkins University, Department of Physics and Astronomy,
Homewood Campus, 3400 North Charles Street, Baltimore, MD 21218-2695;
e-mail: kweaver@milkyway.gsfc.nasa.gov}
\and
\author{Timothy M. Heckman}
\affil{Johns Hopkins University, Department of Physics and Astronomy,
Homewood Campus, 3400 North Charles Street, Baltimore, MD 21218-2695}
\and
\author{Michael Dahlem}
\affil{European Space Research and Technology Centre, Space Science
Department, Astrophysics Division, Postbus 299, NL-2200 AG Noordwijk,
The Netherlands}

\begin{abstract}

We have undertaken an X-ray survey of a far-infrared flux limited 
sample of seven nearby edge-on starburst galaxies.  The data are 
presented in paper 1 of this series by Dahlem, Weaver, \& Heckman.  
Here, we examine the two X-ray-brightest sample members NGC\,253 
and M\,82 in a self-consistent manner, taking account of the spatial 
distribution of the X-ray emission in choosing our spectral models. 
Both galaxies require at least three components to model the 0.1 to 
10.0 keV spectrum, but the modeling is by no means unique.  There is 
significant X-ray absorption in the disk of NGC\,253.  When this is 
accounted for in spectral fits to the emission between $\sim0.2$ and 
2.0 keV, we find that multi-temperature thermal plasma models with 
significant underlying soft X-ray absorption are more consistent with 
the imaging data than single-temperature models with highly subsolar 
abundances or models with minimal absorption and non-equilibrium 
thermal ionization conditions, as have been proposed by others. 
Our three-component models do \underline{not} require absolute 
abundances that are inconsistent with solar values or unusually 
supersolar ratios of the $\alpha$-burning elements with respect to 
Fe (as claimed previously). We conclude that with current data, the 
technique of measuring abundances in starburst galaxies via X-ray 
spectral modeling is highly uncertain.  Much improved spatial/spectral 
resolution and increased sensitivity are required.

Based on the point-like nature of much of the X-ray emission in the 
PSPC hard-band image of NGC\,253, we suggest that a significant 
fraction of the ``extended'' X-ray emission in the $3-10$ keV band 
seen along the disk of the galaxy with {\it ASCA} and {\it BeppoSAX} 
(Cappi et al.) is comprised of discrete sources in the disk, as 
opposed to purely diffuse, hot gas.  If a significant amount of the 
hard X-ray emission were due to unresolved point sources with weak 
Fe K$\alpha$ emission (e.g., X-ray binaries), this could explain the 
low Fe abundances of $\sim 1/4$ solar derived by Cappi et al.\ for 
pure thermal models. 

\end{abstract}




\section{Introduction}

NGC\,253 and M\,82 are the best-studied starburst galaxies in the 
X-ray band.  Their intricate nature has been revealed in observations 
with {\it Rosat} and {\it Einstein}. M\,82 has a complex, central 
star-forming region that contains unresolved point sources, diffuse 
disk emission, and an X-ray ``cone'' pointed along the minor axis of 
the galaxy that is consistent with the outflow of gas in a partly 
confined jet (Bregman, Schulman, \& Tomisaka 1995; Watson, Stanger 
\& Griffiths 1984). NGC\,253 has distinct components within its 
central region (Fabbiano \& Trinchieri 1984) in the form of point 
sources in the disk, diffuse disk emission, unresolved nuclear 
emission, and an extension to the south of the nucleus 
associated with an outflow, visible e.g., in H$\alpha$ emission 
(Watson, Stanger, \& Griffiths 1984).  Both M\,82 and NGC\,253  
possess X-ray emission that extends $6'$ to $9'$ outside the plane of 
the galaxy (Schaaf et al.\ 1989; Fabbiano 1982) with properties 
consistent with a starburst-driven outflow (Dahlem, Weaver,
and Heckman 1998; hereafter paper~1). 

The marked complexity of these objects has resulted in contradictory 
results from instruments that cover different bandpasses and different 
spatial regions. For example, while the hard X-ray detectors onboard 
{\it EXOSAT} and {\it Ginga} measured thermal temperatures of k$T \sim$ 
6 to 9 keV (Ohashi et al.\ 1990, Schaaf et al.\ 1989), soft X-ray 
detectors such as {\it Einstein}, measured significantly lower 
temperatures (Fabbiano 1982).  Recent experiments with a more complete 
energy coverage have at least partly resolved this duality.  Observations
with the Broad Band X-ray Telescope, the Advanced Satellite for 
Cosmology and Astrophysics ({\it ASCA}), and BeppoSAX show that 
multiple spectral components are necessary to model the data between 
0.5 and 10 keV (Petre 1993, Ptak et al.\ 1997 [hereafter P97], Moran 
\& Lehnert 1996 [hereafter ML96], Tsuru et al.\ 1997 [hereafter T97]).
What is not yet clear is whether all of these 
components are required to be thermal in nature (Ohashi et al.\ 1990, 
P97, Cappi et al.\ 1999 [hereafter C99]).

A clear understanding of the X-ray properties of starburst galaxies 
requires observations with a large energy coverage and good spatial 
resolution.  The best such dataset currently available is contained 
in the {\it Rosat} and {\it ASCA} archives. M\,82 and NGC\,253 have 
been the subject of at least six recent papers that focus on the 
archival data (ML96, P97, T97, Strickland et al.\ 1997, Vogler 
and Pietsch 1998 [hereafter VP98], paper 1). Results at higher energies are 
also becoming known from {\it BeppoSAX} (C99).
Based on the innate complexity of M\,82 and NGC\,253, it is not 
surprising that the interpretations of these data by various 
authors differ.  ML96 find a complex absorption structure in      
the core of M\,82 and spectral data consistent with solar
abundance ratios.  On the other hand, P97 and T97 find subsolar 
abundances as well as an overabundance of the $\alpha$-burning 
elements (Mg, Si, S, Ne) with respect to Fe, which suggests that 
Type II supernova remnants dominate the chemical enrichment 
of the X-ray emitting plasma. Since superwinds provide a method for 
enriching the interstellar medium (ISM), these results have 
far-reaching implications for studies of supernova rates and 
outflow models of the enrichment of the ISM. It is thus critical to 
establish how reliable  the current results are.

\section{The Scope of This Investigation}
 
The purpose of this paper, which is part of a larger study
of a sample of seven edge-on starburst galaxies (paper~1),
is to thoroughly examine the 
available X-ray data of NGC\,253 and M\,82 with the goal of
finding a {\it self-consistent spatial and spectral} model
description of the observed X-ray emission between 0.1 and 
10 keV.  Combining the {\it ASCA} and {\it Rosat} datasets is a powerful
spectroscopy tool, because all emission components can be
modeled at one time. The {\it Rosat} spectra, after excision of
point sources, are best-suited to determine the spectral properties
of the $\sim10^6 - 10^7$ K gas emitting 
the softest X-rays, while the hard
X-ray ($> 1-2$ keV) components inferred from these spectra
are subsequently constrained based on the {\it ASCA} fits.
This paper addresses a very fundamental $physical$ problem,
namely the determination of the appropriate composite spectral model 
for starburst galaxies, despite, as we will show, an
ambiguity in the minimum $\chi^2$ space.
Once the most likely physical models are determined, 
similar models can be applied to starburst galaxies with
poorer quality data (for example, fainter galaxies or those at  
high redshift). 
 
\section{Spectral Analysis}

\subsection{A Note on the Spectral Modeling}

\begin{figure}
\epsfig{file=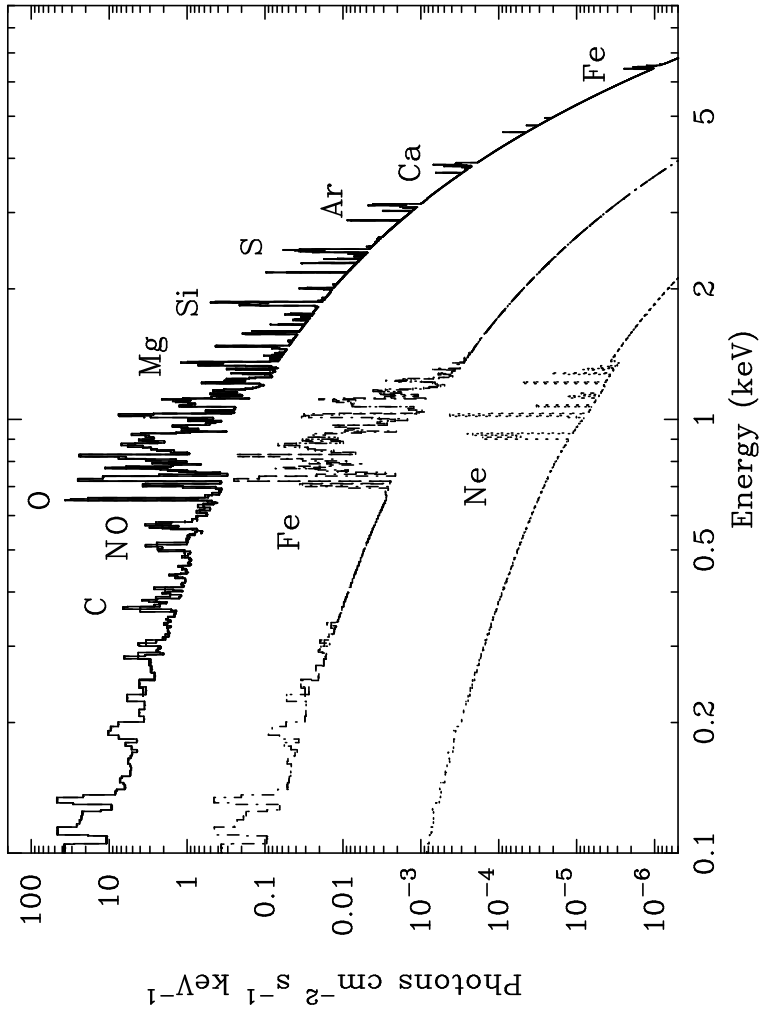, height=15cm, rotate=90}
\caption[ ]{MEKAL plasma model for a k$T$ = 0.5 keV 
solar-abundance plasma (top curve).  The strongest emission lines 
from each element are labeled with the exception 
of the Fe and Ne complexes, which are shown separately for clarity in 
the middle and bottom curves.  The Y axis is arbitrary.  
\label{fig:mekal}
}
\end{figure}

There is strong evidence that most of the soft X-ray emission in NGC\,253
and M\,82 arises from thermal processes.  
To model the soft X-ray emission ($< 2$ keV) we consider Raymond-Smith 
plasma models (RS; Raymond \& Smith 1991) and Mewe-Kaastra plasma 
models with Fe L calculations by Liedahl (MEKAL; Mewe, Gronenschild 
\& van den Oord [1985], Mewe, Lemen \& van den Oord [1986], Kaastra 
[1992]).  Typical thermal temperatures might be similar to those 
found for diffuse gas in normal galaxies, which fall near a few 
tenths of a keV.  The X-ray spectrum from such a plasma is rich in 
line emission near 1 keV (Figure~\ref{fig:mekal}). Current data are 
inconclusive as to whether the hard X-ray emission is of thermal 
origin (Ohashi et al.\ 1990, P97, T97, but see C99).  The {\it ASCA} data 
are not particularly sensitive to the Fe K line emission expected for 
a thermal plasma (P97), and so for simplicity, we adopt bremsstrahlung 
or power law models to describe the $\sim$2--10 keV continuum.

Our results are tabulated in Tables~\ref{tab:n253_pspcfits} --
\ref{tab:m82jointflux}.  Each fit is assigned one 
or a combination of the following 
codes for the model used, together with the number of the fit if 
more than one fit is listed.  The models consist of:

\begin{itemize}

\item ``P'' -- a power law,

\item ``B'' -- a thermal Bremsstrahlung spectrum,

\item ``M'' -- a MEKAL plasma model,

\item ``R'' -- a RS plasma model.

\end{itemize}

All models include absorption due to our Galaxy and additional 
absorption where necessary.

\subsection{The Definition of Different Emission Components}
\label{par:pspcregions}

\begin{figure}
\epsfig{file=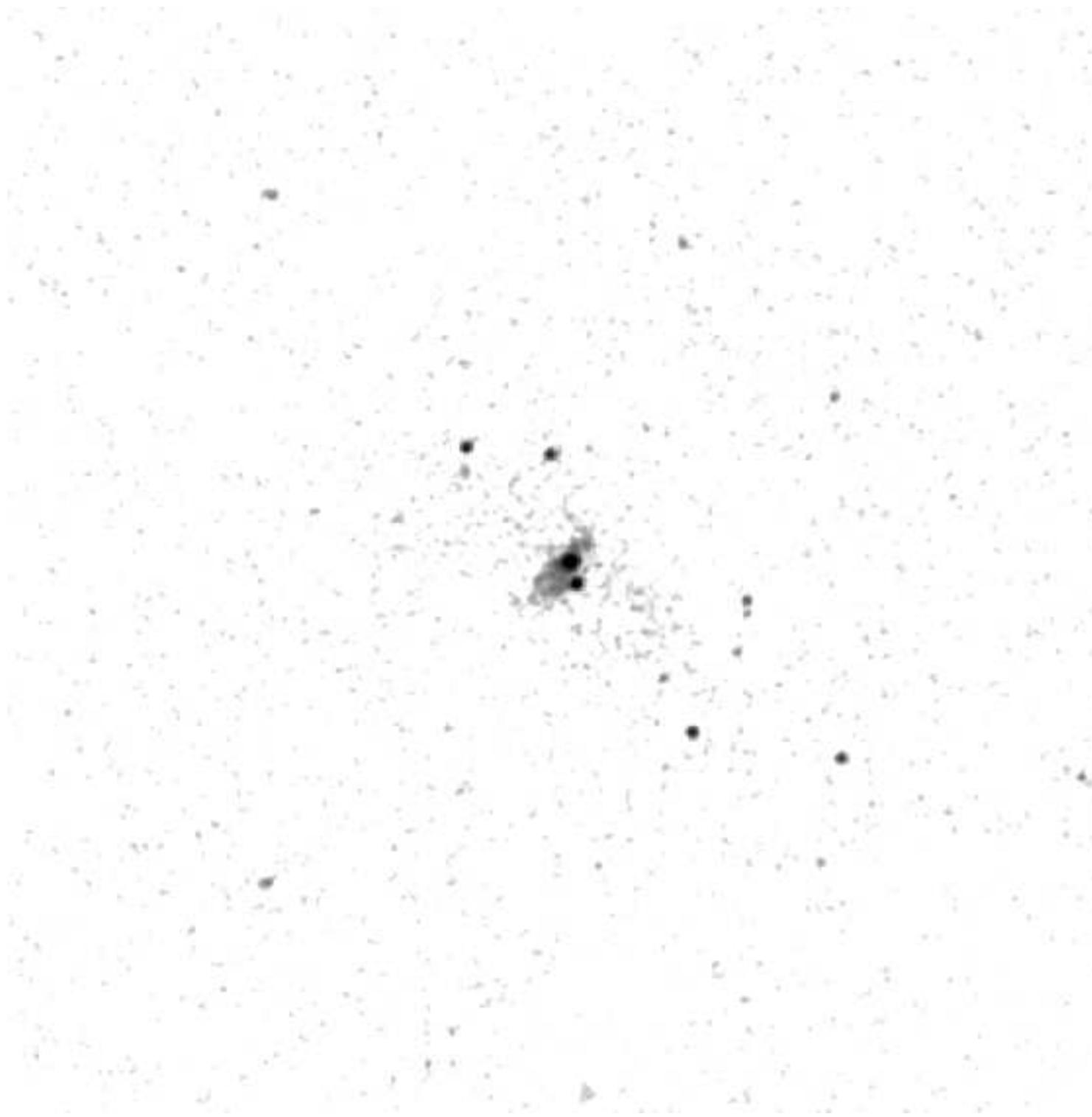}
\caption[ ]{The {\it Rosat} HRI image of NGC\,253.  The field 
is $25' \times 25'$ and the data are smoothed for an effective 
resolution of $10\farcs5$ (cf. paper~1). 
\label{fig:n253_hri}
}
\end{figure}

\begin{figure}
\epsfig{file=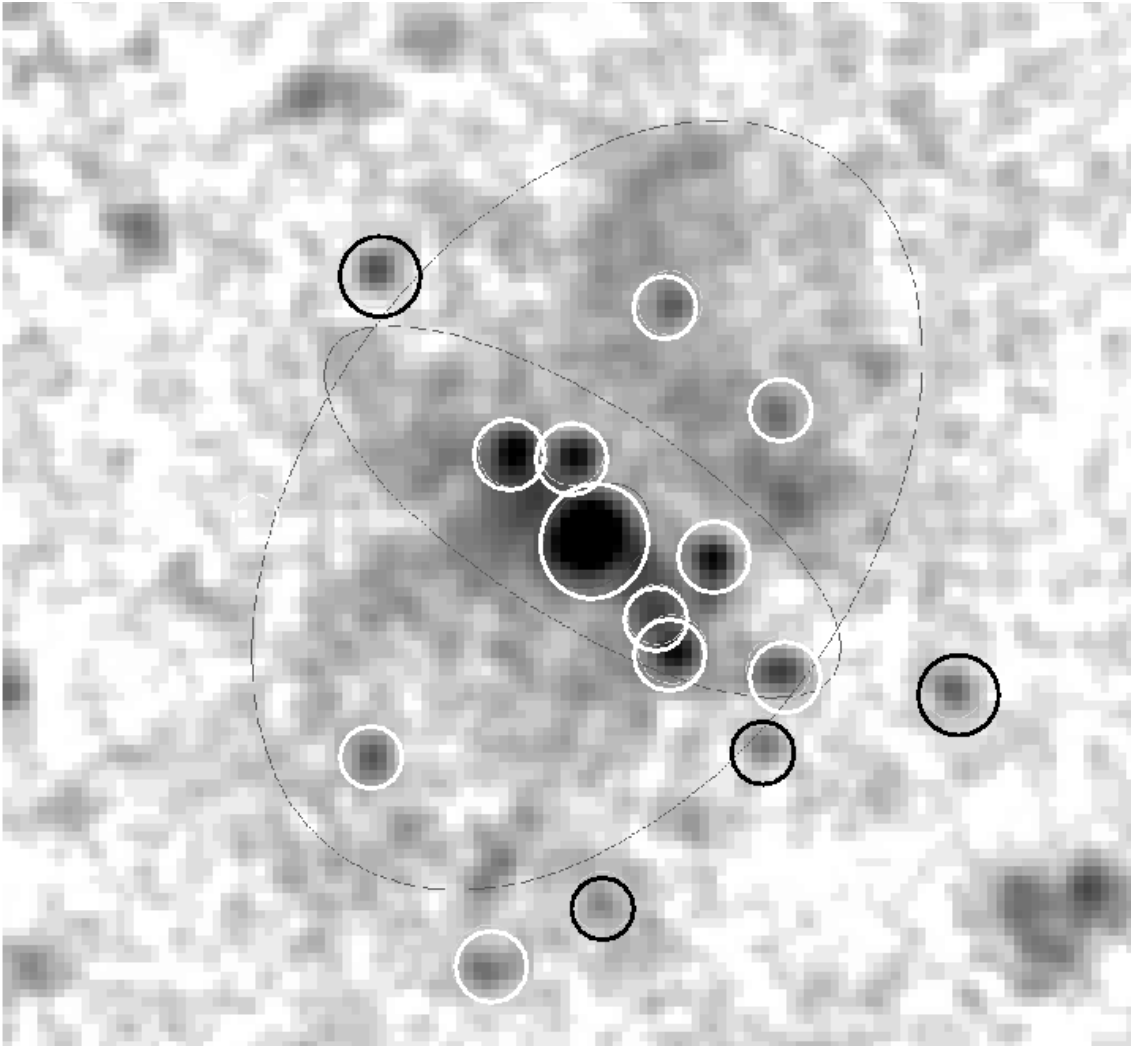, width=15cm}
\caption[ ]{PSPC image binned to have $15''$ pixels and   
consisting of photons that fall within the 0.25 keV band and 
1.50 keV band as defined in paper~1.  The circular and elliptical 
regions used to extract the spatially-resolved spectra of the 
galaxy halo, disk and point sources are drawn.
\label{fig:n253_regions}
}
\end{figure}

Our imaging analysis and techniques for spectral
extraction are discussed in paper~1.
The emission components within each galaxy
are spatially identified in the {\it Rosat} HRI and PSPC 
images as follows:

\begin{enumerate}

\item NGC\,253:

At a limiting flux level of $\sim6\times10^{-14}$ ergs cm$^{-2}$ 
s$^{-1}$, we identify eighteen point sources in the HRI image 
(Figure~\ref{fig:n253_hri}) within a $\sim12'$ radius of the galaxy 
core.  Based on the smoothed and point-source-subtracted PSPC images 
(paper~1), we divide the galaxy into ``halo'', ``disk'', and ``core'' 
regions (Figure~\ref{fig:n253_regions}). To examine the band of 
absorption apparent in the 0.25 keV PSPC image (paper~1, Figure 12), 
we also divide the elliptical region that delineates the disk equally 
into northern and southern regions (each 1/2 of the NE-SW ellipse in 
Figure~\ref{fig:n253_regions}).  Compact sources are excised and 
classified as ``hard'' or ``soft'' by comparing counts in the 0.25, 
0.75, and 1.5 keV bands. Their spectra are then 
co-added\footnote{The hard compact sources are NGC\,253:DWH 1, 2, 
4, 6, 9, 12, 13, 14, and 15 in the PSPC image; the soft sources are 
NGC\,253:DWH 3, 16, and 17 (paper~1).}
(\S~\ref{par:pspc253spec}).  

{\it ASCA} observed NGC\,253 with the SIS operating in 4-CCD mode, 
which provides a $22' \times 22'$ field of view.  NGC\,253 is 
approximately $25' \times 12'$ in size and so the galaxy is contained 
almost entirely on two of the four SIS chips, with only about 15\% 
of the halo flux lost off the edges of the chips. For our joint PSPC 
and {\it ASCA} analysis, spectra are extracted from 
$10\farcm5\times15\farcm0$ rectangular regions in the SIS and 
15$^{\prime}$ diameter circular regions in the GIS.  This technique 
includes as much of the halo flux as possible. For a direct comparison 
of our {\it ASCA} results with those of P97 (\S~\ref{par:asca253spec}), 
we alternatively use circular regions identical to theirs (10$^{\prime}$ 
in diameter).  SIS background is accumulated from the non-source chips.

\item M\,82:

Based on the {\it Rosat} images, we divide the galaxy into halo and
core regions.  Of the 17 point-like sources identified in the
PSPC image (paper~1, Figure 13), none are located within the galaxy.  
These sources are sorted by hardness ratio and co-added as above for 
spectral fits\footnote{The soft compact sources are M\,82:DWH 1, 2, 8, 
9, 10 and 15; all others have hard spectra (paper~1).}.

\end{enumerate}

\section{Spectroscopic Results for NGC\,253}
\label{par:n253spec}

\subsection{Modeling the PSPC Spectra}
\label{par:pspc253spec}

From our PSPC imaging analysis (paper~1), we label the predominant 
spatial/spectral sources as (1) the diffuse gaseous halo, which we 
identify from the images with emission at low X-ray energies (0.25 
keV band), (2) the galaxy disk, which we identify from the images 
primarily with emission at medium X-ray energies (0.75 keV band), 
and point sources (including the core), which we identify from the 
images with emission at high X-ray energies (1.5 keV band).

\begin{figure*}
\epsfig{file=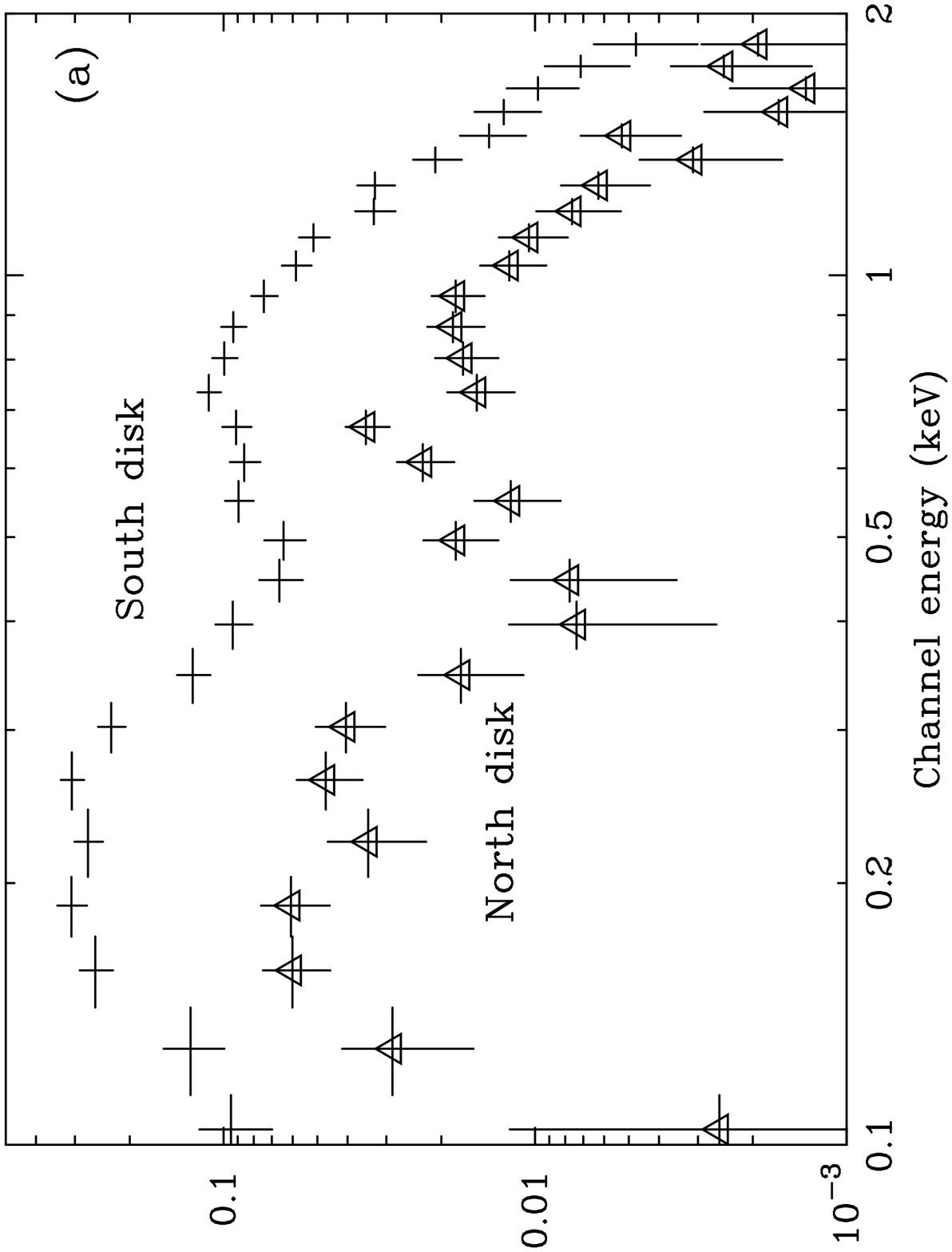, width=6.5cm}
\hspace*{1cm}
\epsfig{file=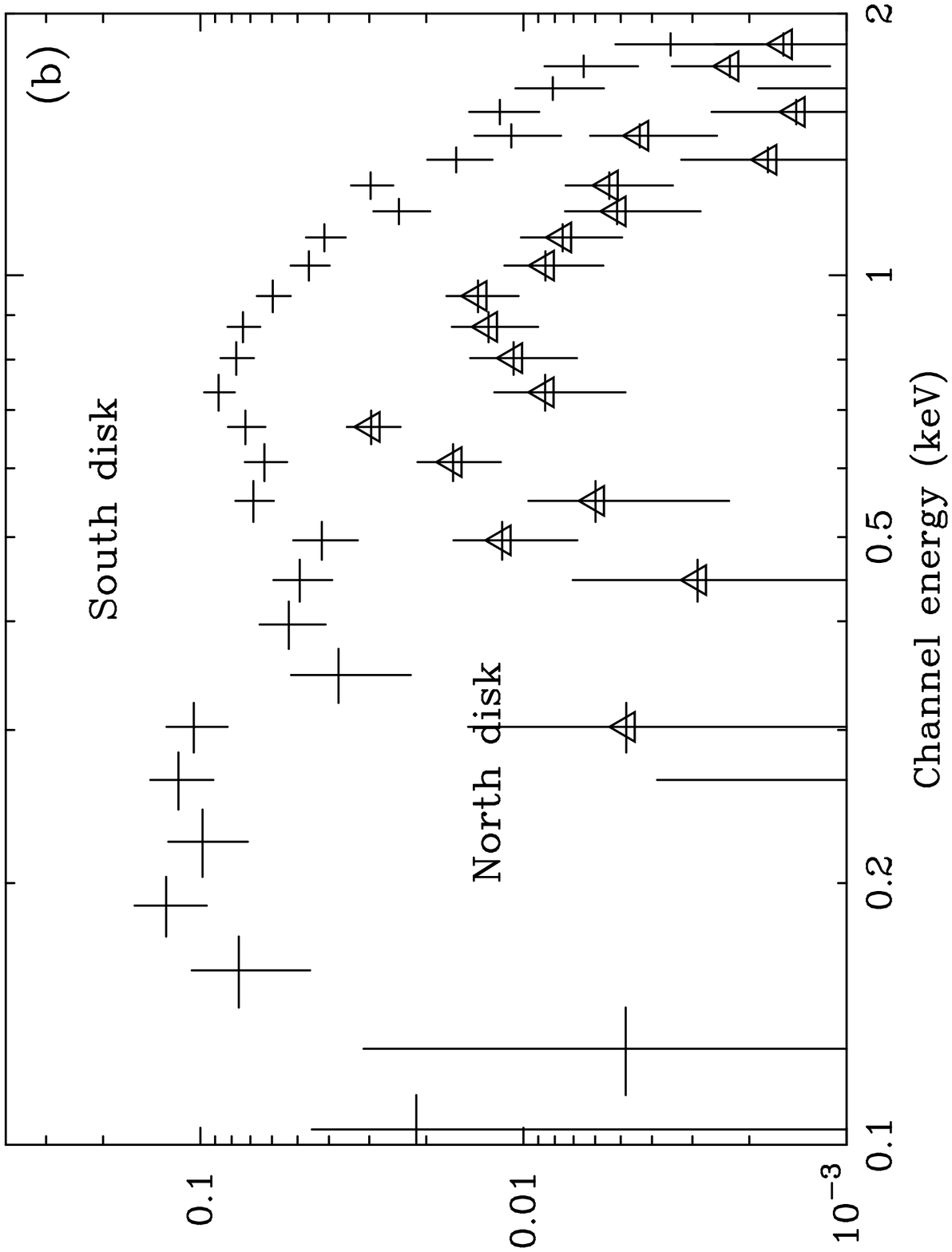, width=6.5cm}
\caption[ ]{PSPC spatially-resolved spectra of the diffuse 
disk emission with compact sources removed.  For (a) the field 
background is subtracted and for (b) the 
local background is subtracted.  Crosses represent the southern 
disk and triangles represent the northern disk.  
\label{fig:disk_spectra}
}
\end{figure*}

\begin{enumerate}

\item The Diffuse Halo Emission:

To describe the halo emission we examine one- and two-temperature 
thermal models.  The simplest case is a single-temperature MEKAL 
plasma model with Galactic absorption (\NH\ = 0.9$\times$10$^{20}$ 
cm$^{-2}$) and solar abundances.  Such a model provides a very poor 
fit to the data ($\chi^2 = 96.1$ for 28 \dof, model M1, 
Table~\ref{tab:n253_pspcfits}).  When the  
abundance is a free parameter, the fit improves dramatically 
($\chi^2$ = 20.1 for 27 \dof, model M2), but the implied abundance  
of $Z=0.01$ \zsol\ is extremely low.  We cannot be sure that this 
model provides a correct {\it physical} description of the data. 
Indeed, when we consider the {\it ASCA} data, our fits suggest 
generally higher abundances for thermal models. Given the limited 
bandpass and poor energy resolution of the PSPC, it is possible that
continuum flux from a warmer gas can contribute to mimic abundances 
that are less than predicted in the cooler gas.

When the data are fitted with a two-component thermal plasma model, 
the abundances can be as large as solar but with a fairly large 
uncertainty. If we assume an abundance of $Z=0.2$ \zsol\
(Table~\ref{tab:n253_pspcfits}, model MM) then 
k$T_1 = 0.14^{+0.03}_{-0.06}$ keV and k$T_2$ = 0.61$\pm$0.36 keV 
($\chi^2$=19.6 for 26 \dof). The observed and absorption-corrected 
$0.1-2.0$ keV fluxes for this case are 1.24 and $1.89\times10^{-12}$ 
ergs cm$^{-2}$ s$^{-1}$, respectively, with relative contributions 
of 40\% and 60\% from the cool and warm components.  This ratio is 
consistent with the ratio of extended halo emission in the 0.25 keV
and 0.75 keV PSPC maps (paper~1), and so we are confident that our 
two-component model is appropriate even though there is a large 
uncertainty on the abundance.
 
\item The Diffuse Disk Emission:

For the disk emission, a power-law (model P) or a MEKAL plasma (model 
M) provide reasonable fits.\footnote{When dealing with the inner 
regions of the galaxy, the choice of background is important.  Poorer 
fits result when field background is subtracted, and  even though we 
do not know the geometry of the system or how much of the diffuse halo
emission is projected onto the disk {\it a priori}, the fits using 
local background are more reasonable and better match the imaging 
data.  Thus we rely on the latter method.} The large photon index of 
$\Gamma$=3.68 for model P (a steep spectrum) implies a significant 
thermal plasma contribution, but model M again implies an unusually 
low abundance of $Z$ = 0.003 \zsol. As for the halo fits, we argue 
that such a low apparent abundance can result from a model that is 
too simple. 
To find a more physically reasonable description of the data we
examine more complex models choosing {\it a priori} values of 
$Z$ = 1.0 \zsol\ and $Z$ = 0.5 \zsol\ for the abundance.  Both 
are consistent with the data.

For solar and one-half solar abundances, the data are well described 
with a model that consists of a power-law and MEKAL component (Model 
PM1, Table~\ref{tab:n253_pspcfits}).  For this model the best-fitting 
power-law index is $\Gamma = 0.77$; however when the {\it ASCA} data 
are taken into consideration, if $\Gamma$ were really this small
it would seriously overpredict the flux above 2 keV when extrapolated 
to higher energies (\S~\ref{par:asca253spec}).  Also, if the ``flat'' 
hard component were thermal, the required temperature would be much 
too high and also inconsistent with the {\it ASCA} data. Alternatively, 
we suggest that the small index is an artifact of the effect of 
absorption on an intrinsically steeper hard X-ray component.  If we allow 
for some additional absorption and fix $\Gamma$ at the best-fitting 
{\it ASCA} value of 1.9 (Model PM2, Table~\ref{tab:n253_pspcfits}), 
we derive a column density of $\sim10^{21}$ cm$^{-2}$.  This 
model is more reasonable than the model with the unusually small 
power-law index and it is also consistent with the images since we 
see the absorption band due to the galaxy disk cutting in at around 0.7 keV.
  
To confirm the significant absorption in the galaxy disk we examine 
the northern disk (ND) and southern disk (SD) separately.  Both 
spectra are plotted in Figure~\ref{fig:disk_spectra}.  Subtracting  
field background, the spectra have a similar shape even though there 
is a clear band of absorption in the image, which confirms our 
suspicion that diffuse halo emission is projected {\it onto} the 
band of underlying absorption in the disk.  When the local (diffuse) 
background is subtracted  \NH(ND) is 2 to 3 times \NH(SD), which is 
more in line with the image. Depending on the choice of models 
(power law or MEKAL), the intrinsic absorption for the hard component, 
i.e., due to the disk, can be as large as $6 \times 10^{21}$ cm$^{-2}$.

\item The Core Emission:

Single-component power-law or MEKAL models provide poor fits to the 
core spectrum (models P and M, Table~\ref{tab:n253_pspcfits}). On the 
other hand, a two-component MEKAL plasma ($Z \approx 0.5$ \zsol) plus 
power-law model provides an excellent fit with $\chi^2$/$\nu$ = 27.6/24
(Table~\ref{tab:n253_pspcfits}, model PM1). In this case, $\Gamma$(PSPC) 
is smaller than $\Gamma$({\it ASCA}), a result similar to the hard model 
component of the disk emission.  The small index again implies significant 
absorption. Fixing $\Gamma$ at the {\it ASCA} value of 1.9 (model PM2) 
yields \NH\ = 5 $\times 10^{20}$ cm$^{-2}$.  Constraining the hard 
component in this way also causes the fitted temperature of the thermal 
component in the PSPC to be equal to the temperature derived from 
{\it ASCA}.  This provides a nice consistency check and allows us 
to identify the soft component in the core spectrum with the warm 
{\it ASCA} thermal component (P97). If the hard component is {\it more} 
absorbed than the thermal component (model PM3, our best fit), then 
\NH(core) can be as high as 4 to $5 \times 10^{21}$ cm$^{-2}$.  This 
confirms the amount of absorption intrinsic to the galaxy disk and 
is consistent with that inferred by VP98.

\item Compact Sources:

The compact sources in NGC\,253 are discussed at length by VP98, 
who perform a sensitive search for point sources in the disk,
resulting in 27 detections in the HRI image. We do not perform a 
similarly detailed analysis because our intent is only to identify 
the major point-source contributions in the PSPC data
(Fig.~\ref{fig:n253_regions} and paper~1). Our integrated PSPC 
point-source luminosity is $\sim2\times10^{39}$ ergs s$^{-1}$, 
which is comparable to that of VP98, so we are confident that we have 
identified the majority of the flux from the point sources.  
Anything else is not important for our investigation.
  
The ``hard'' X-ray point sources can be modeled with a power law 
having $\Gamma \sim 2$ (model P, Table~\ref{tab:n253_pspcfits})
or a bremsstrahlung spectrum with kT $\sim1.9$ (model B).  
The index of the power law is 
similar to $\Gamma$({\it ASCA}) for the hard component and implies that the 
{\it Rosat}-identified point sources comprise a significant portion 
of the hard X-ray portion of the {\it ASCA} 
spectrum. Our result suggests that much of the 
hard X-ray emission is truly point-like, in contrast to the claim 
of C99 that the $3-10$ keV emission is primarily from hot, diffuse 
gas in the galaxy disk.  From extrapolation, the 
compact sources, including the core, 
provide $30-50\%$ of the total $2-10$ keV emission seen with 
{\it ASCA} and so must make up a significant fraction of the $3-10$ 
keV ``extent'' seen with {\it BeppoSAX} (C99).

The ``soft'' X-ray sources that are point-like in the HRI and the PSPC 
are located exclusively in the halo and have spectral characteristics 
resembling the diffuse gas (Table~\ref{tab:n253_pspcfits}). This 
suggests that the point-like regions in the halo are not distinct 
compact sources but might be areas of locally enhanced halo emission, 
as also suggested by VP98.  For the same two-temperature plasma model 
applied to the halo emission (model MM), 
the cool and warm components contribute 
about 40\% and 60\% of the flux, respectively, similar to the ratio of 
these  components in the halo.

\end{enumerate}

We conclude from our analysis of the spatially-resolved PSPC spectra 
that there is significant absorption within the galaxy disk of 
NGC\,253 on the order of 1 to a few $\times 10^{21}$ cm$^{-2}$ (in 
agreement with VP98). We find the PSPC point sources to be a significant
source of X-rays at 2 keV and they are likely to contribute $30-50\%$ of
the total flux in the $2-10$ keV band, i.e., in the {\it ASCA} and 
{\it BeppoSAX} spectra and the {\it BeppoSAX} image reported by C99.

\subsection{Comparing PSPC and {\it ASCA} Spectral Modeling}
\label{par:asca253spec}

Here we discuss the method of spatial-spectral iteration by which 
we derived a self-consistent modeling of the {\it ASCA} and PSPC 
spectra.  In the above analysis, we used simple models to fit the 
spatially resolved PSPC spectra in our attempt to infer the {\it 
spatial} origins of the {\it ASCA} emission.  The basic assumption 
is that if a PSPC spatial component is mostly hard or mostly soft, 
then this region can be associated with the corresponding hard or 
soft component in the {\it ASCA} spectrum.  Conversely, if we do 
not find a PSPC spatial/spectral component that ``matches'' one
of the {\it ASCA} spectral components, then we know that at least 
one of our model assumptions is wrong.

The {\it ASCA} spectrum clearly contains strong line emission from 
O, Ne, Fe, Mg, Si, S, and Ar (P97 and others).  The integral 
spectrum can be well fitted with the sum of an absorbed hard X-ray 
component (power law or bremsstrahlung) and one or more thermal 
plasma components.  But this is not necessarily the correct model.  
The data are ambiguous in that the amount of absorption for the 
various components and the abundance of the gas can differ by orders 
of magnitude depending on the input assumptions.  

\begin{figure*}
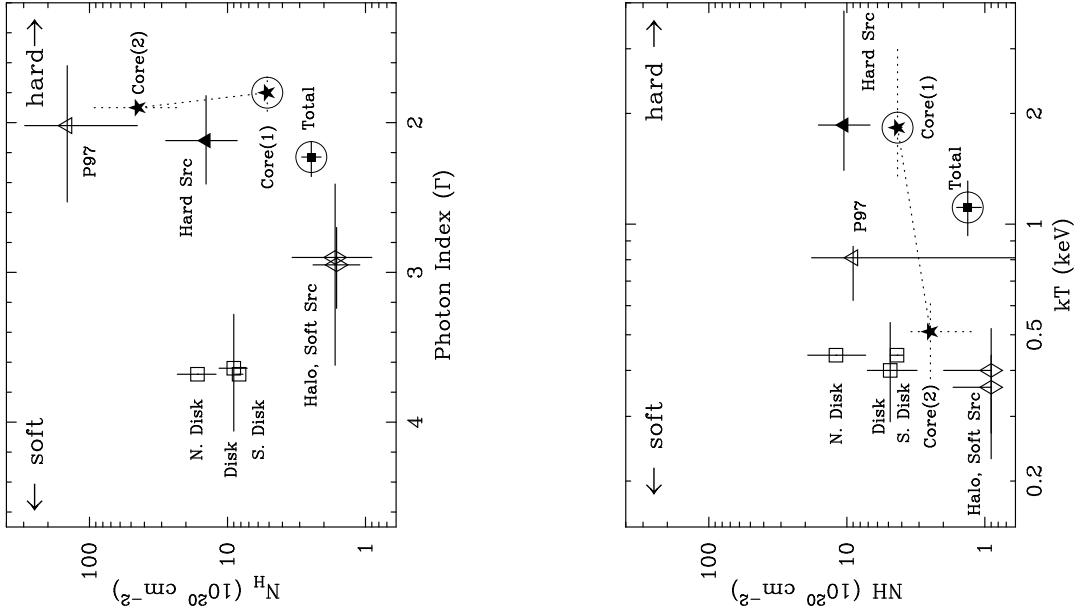

\hspace*{0.8cm}
\epsfig{file=f5a.ps, width=6cm}
\hspace*{2cm}
\epsfig{file=f5b.ps, width=6cm}
\caption[ ]{Model-fitting results for NGC\,253.  
Open squares (disk emission), diamonds (halo, soft point sources), 
stars (core) and 
solid triangles (hard point sources) are the  
spatially-resolved PSPC results for single-component model fits
(either a power law or a MEKAL plasma), except for core(2), which 
results from a two-component power law plus MEKAL plasma model 
fit to the core spectrum.  Open triangles are the {\it ASCA} result 
from the corresponding spectral component from P97.
Circles indicate that a fit is statistically
unacceptable as defined by $\chi^2_{\nu}$ greater than 1.3. (a)
\NH\ vs.\ $\Gamma$ for a power-law model.  The point 
labeled P97 represents the corresponding hard power-law component 
from P97; core(2) represents the hard core component from the 
PSPC (model PM3, Table 1).  (b) \NH\ vs.\ k$T$ for a single MEKAL model 
with free abundances.  The point labeled P97 represents the 
corresponding thermal component from P97; core(2) represents the soft
core component from the PSPC (model PM3, Table 1). 
For comparison we also plot the
single-component model result for the integral PSPC 
spectrum (labeled ``Total'').  Spectra of the  
halo and soft compact sources have field background subtracted;
those for the disk have halo background subtracted; those for
the core have halo plus diffuse disk emission subtracted.  
\label{fig:pspcfits}
}
\end{figure*}

Figure~\ref{fig:pspcfits} is a guide to show which component of the 
X-ray emission, hard or soft, dominates the spatial components in the 
PSPC and hence which {\it ASCA} and PSPC models yield consistent results.
Since our goal is to match up {\it ASCA} and the PSPC, we choose the two
most significant {\it ASCA} components from P97 and compare their spectral  
parameters with the PSPC.  Figure~\ref{fig:pspcfits}a shows $\Gamma$ vs. 
\NH\ for the power-law (hard X-ray) {\it ASCA} component compared with the  
power-law results for the spatially-resolved PSPC spectra.  All PSPC 
points are derived for a single power-law model except for core(2), which 
represents the photon index for the best-fitting 
two-component core model (model PM3, 
Table~\ref{tab:n253_pspcfits}).  The hard X-ray point sources and the 
hard component in the PSPC core clearly have the spectra that are most similar 
to {\it ASCA}.  This consistency provides independent support for our 
complex modeling of the core spectrum. 

Figure~\ref{fig:pspcfits}b shows 
k$T$ vs.\ \NH\ for a MEKAL fit to the {\it ASCA} soft component (P97) and 
single-component MEKAL fits to the PSPC, except for core(2) as above. 
The disk emission with k$T \sim 0.8$ keV most closely corresponds to the
soft {\it ASCA} thermal component found by P97; however, in this case, no 
spatially resolved PSPC component has exactly the same model parameters.  In 
fact, much of the spatially resolved emission has a significantly lower
temperature than measured with {\it ASCA}. These results strongly imply 
a three-component modeling of the PSPC {\it and} {\it ASCA} data, 
including a very soft thermal component that contributes in the {\it ASCA} 
band.  For comparison we also plot the single-component model result for 
the integral PSPC spectrum, which illustrates the incorrect 
results we might infer if the data had poor enough statistics that a 
single-component model were statistically acceptable. 

\subsection{Modeling the {\it ASCA} Spectrum}

We compare our results for two- and three-component  models for {\it ASCA} 
in Table~\ref{tab:n253asca}. To account for systematic differences between 
our fitting techniques and those of P97, we have re-extracted the {\it ASCA}
spectrum using their region sizes. The abundances of individual elements 
are not well determined so we follow P97 and group the plasma abundances
as follows: Ne and Fe (hereafter \znf ), O, Mg, Si, and S (hereafter 
\zomss), and N, Ar, Ca, and Ni (hereafter \zhcn).  In reality, Ne should 
be tied to S, Si, and Mg, based on the physics of production in Type II 
supernovae; however, the fact that the Ne emission is blended with Fe L 
(Figure~\ref{fig:mekal}), combined with the poor energy resolution of the 
detectors, forces us to tie this parameter to Fe. This leaves 13 free 
parameters for the {\it ASCA} fits, which are the normalization of the 
cool component\footnote{We do not measure multiple abundances for the 
cool component because the line emission for the $\sim$0.2 keV plasma is 
dominated by Ne and Fe L and the lines at higher energies are swamped 
by the emission from the warmer gas.}, the temperature (k$T$(w)), 
normalization, \zomss, \znf, \zhcn, and the absorbing column (\NH(w)) 
of the warm component, the index (or temperature), normalization, and 
absorbing column (\NH(h)) of the hard (or ``hot'') component, and 
the relative normalizations of the instruments. The Galactic foreground 
column density of 0.9$\times$10$^{20}$ cm$^{-2}$ is included in all cases.  
Based on joint PSPC and {\it ASCA} fits (\S~\ref{par:joint253spec}) we 
choose \znf,\zhcn\ = 1.0 \zsol\ and fix the temperature and abundance 
of the cool gas at k$T$(c) = 0.2 keV and $Z = 0.2$ \zsol, respectively.

\begin{figure*}
\hspace*{2cm}
\epsfig{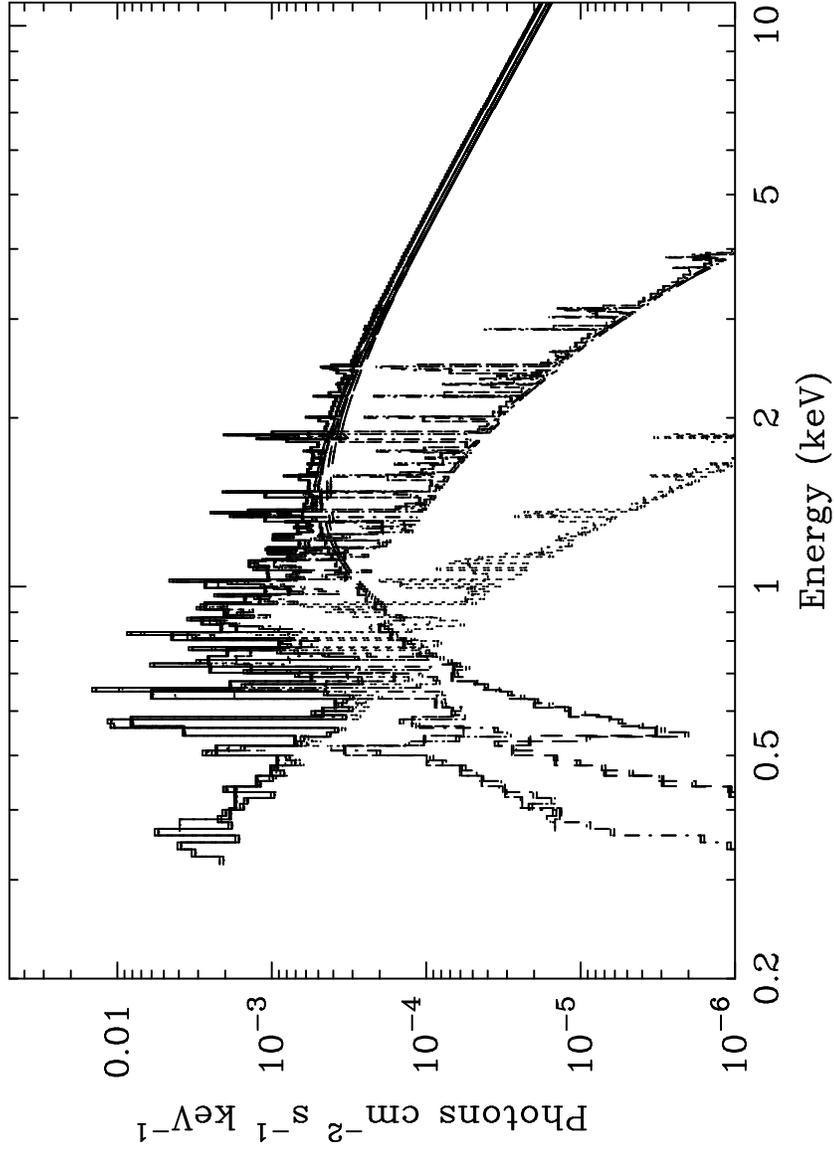}
\caption[ ]{Our best fitting {\it ASCA} model for NGC\,253 which 
consists of a two-temperature MEKAL plasma with temperatures 
of k$T$ = 0.2 keV and  k$T$ = 0.6 keV and a power law with 
$\Gamma$=2.0.   The abundances are fixed at 0.2 \zsol\ for the 
cool component and 1.0 \zsol\ for all but O, Mg, Si, and S for 
the warm component. The derived absorption is $4\times10^{21}$ 
and $84\times10^{21}$ cm$^{-2}$ for the warm plasma and power-law 
components, respectively; see Table~\ref{tab:n253asca} (model PMM).
\label{fig:asca253spec}
}
\end{figure*}

We derive results similar to P97 for a two-component model 
(models PM or BM in Table~\ref{tab:n253asca}) but find 
equally good or better fits for a three-component model
(models PMM or BMM).  We conclude 
that the three-component and two-component models are statistically 
equivalent (the intrinsic uncertainty in $\chi^2_{\rm min}$ results 
from the choice of plasma models) and ambiguous.  But when we account 
for the low-temperature component in the PSPC, the three-component 
model is much more consistent with the imaging data. The three-component 
model not only allows the metal abundances of the warm component to be 
$1-2$ \zsol (which is physically more meaningful for a galaxy with a 
very high star formation rate, cf. Zaritsky, Kennicutt and Huchra 1994), 
it also allows \NH(h) to be smaller and more consistent with \NH\ 
inferred from PSPC fits.  Our best-fitting {\it ASCA} model (PMM, 
Table~\ref{tab:n253asca}) is shown in Figure~\ref{fig:asca253spec}.

\subsection{Joint Spectral Modeling}
\label{par:joint253spec}

\begin{figure}
\epsfig{file=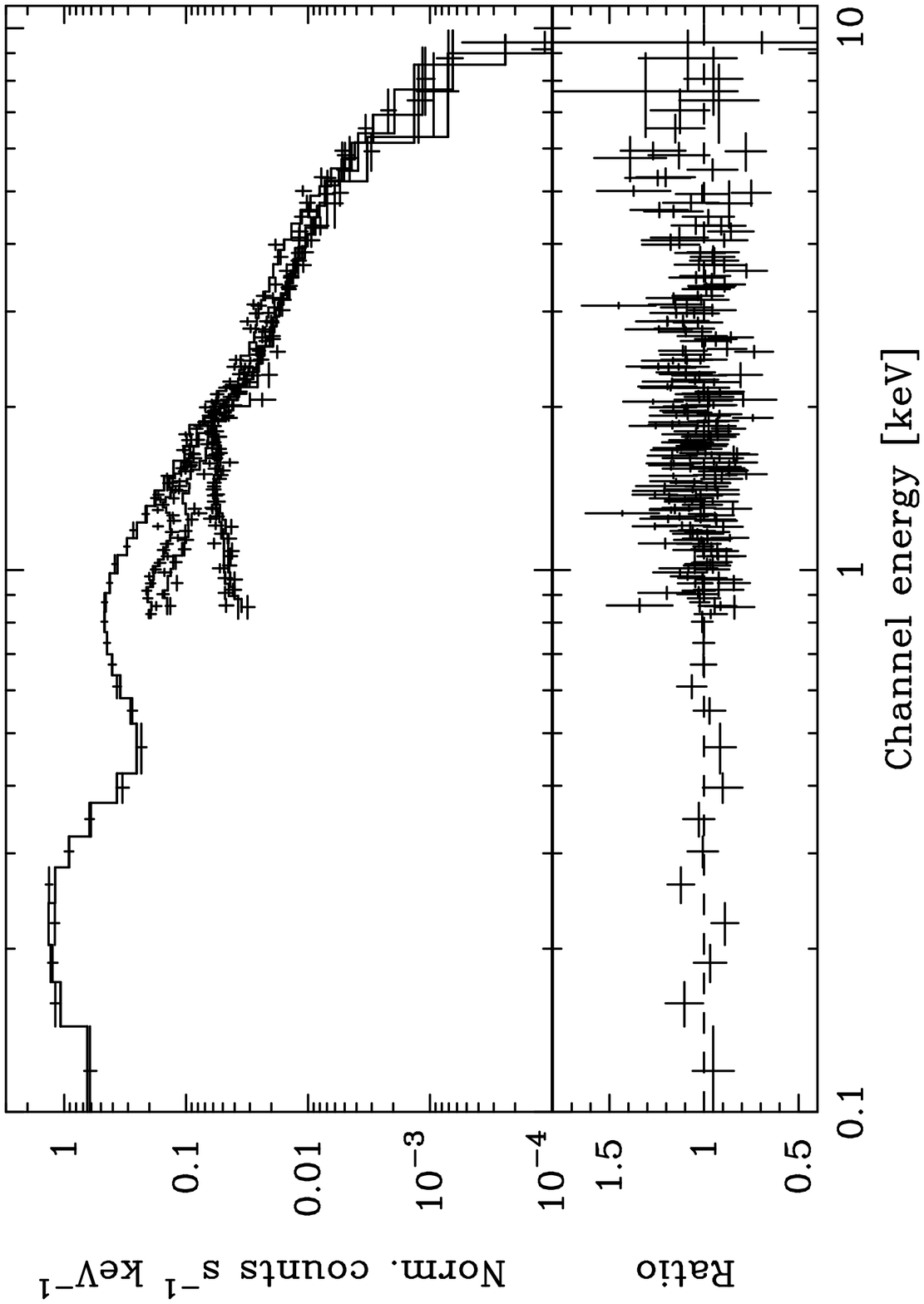, width=6.2cm}
\hspace*{1cm}
\epsfig{file=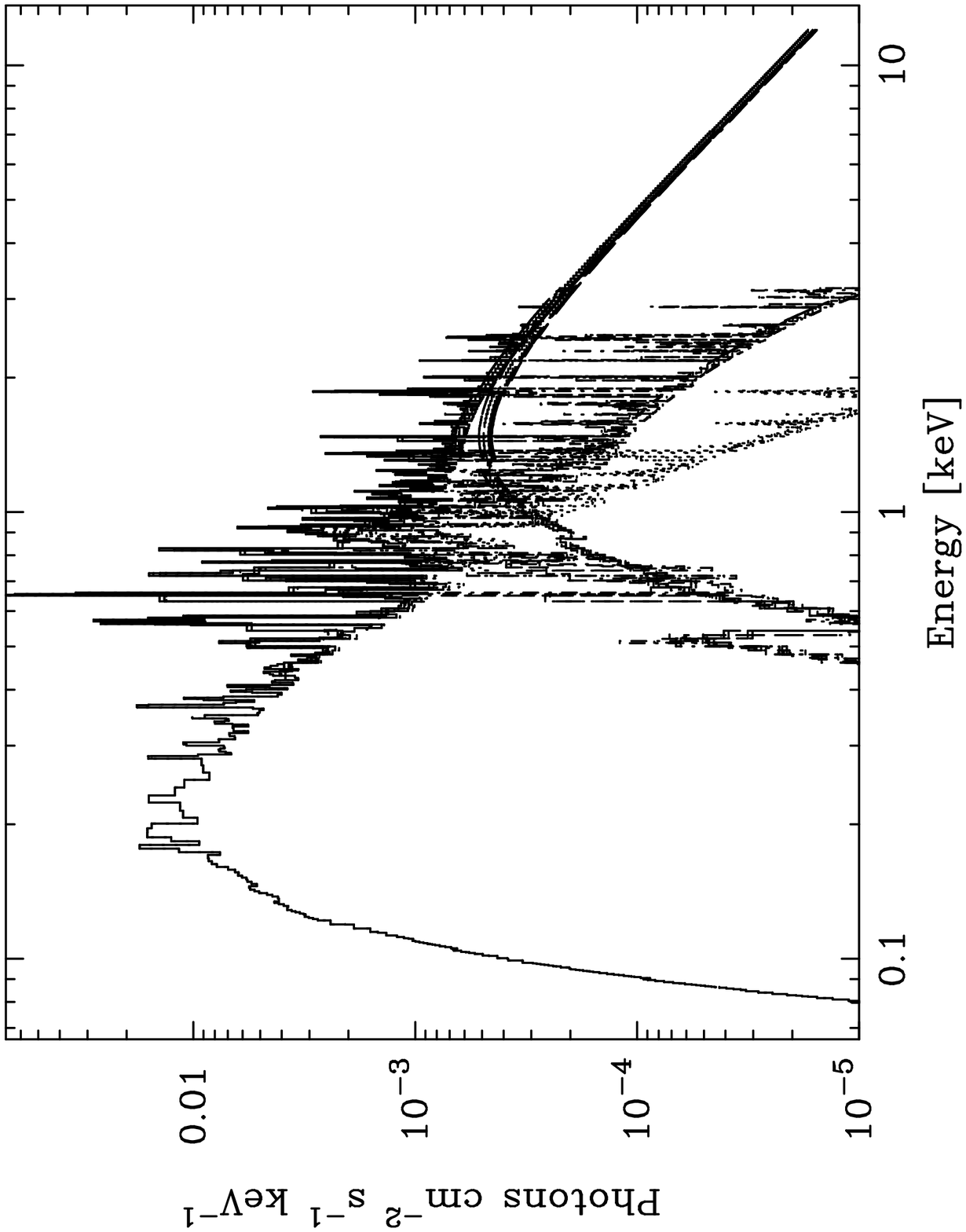, width=6.2cm}
\caption[ ]{(left) {\it ASCA} and {\it Rosat} data for NGC\,253 and 
the best-fitting two-temperature MEKAL plus power-law model folded through the 
instrumental response (Table~\ref{tab:n253joint}, $\chi^2$=540.5).  
The ratio of the data to this model is plotted in the bottom panel.
(right) The model (Table~\ref{tab:n253joint} model PMM2).
\label{fig:joint253spec}
}
\end{figure}

Having demonstrated the large uncertainty in deriving absolute 
abundances for thermal models, 
we fix \zhcn\ at values of 0.5 \zsol\ or 1.0 \zsol for the  
joint {\it ASCA} and PSPC fits.  With 
\zhcn\ = 0.5 \zsol\ (Table~\ref{tab:n253joint}), the best fit is 
obtained with two-temperature MEKAL plasma plus a power law ($\chi^2$ 
= 538.5 for 499 \dof, model PMM1). For this model we derive \zomss\ = 
0.8 and \znf\ = 0.4. However, $\chi^2$ space is shallow in the region 
near $\chi^2_{\rm min}$ due to \znf, \zomss, $N_{\rm H}$(h), and 
$N_{\rm H}$(m) being strongly coupled. After thorough testing of 
the allowed parameter space, we conclude that there is no need for 
the Fe abundances to be less than solar. The joint data and 
best-fitting two-temperature MEKAL plasma plus power law model with 
\znf\ = 0.5 \zsol\ are shown in Figure~\ref{fig:joint253spec}
(model PMM2).

In general, $\Gamma$, k$T$(c), k$T$(w), and k$T$(h) are well 
determined for the joint fits, while there are large uncertainties 
in \NH(w), \NH(h) and \zomss.  \zomss\ tends to be larger than \znf, 
but there are cases where the two are consistent to within the 
statistical errors.

The mean value for \NH(w) is 5.81 $\times$ 10$^{21}$ cm$^{-2}$ and
the mean value for \NH(h) is 6.72 $\times$ 10$^{21}$ cm$^{-2}$.
Both are consistent with \NH(core) inferred from the PSPC, but this 
is slightly larger than we might expect for \NH(w).  If we add a 
third MEKAL plasma component with k$T$ = 0.14 keV to match the softest 
emission in the PSPC, then \NH(w) can be lower and more consistent 
with the PSPC data.  This also allows the abundance in the halo to 
be higher than 0.2 \zsol. An alternative explanation of an effectively 
large \NH(w) is a more complex absorption structure.  If the hard X-ray 
component is modeled as being partially covered, about 30\% of the hard 
X-ray flux could be unabsorbed or absorbed by a very low column
[model (PMM)pc, Table~\ref{tab:n253joint}].
In other words, if some of the hard 
X-rays are passing through a patchy absorber, then it would {\it add} 
flux at low energies, another reason why we might measure low abundances 
below 1 keV for purely thermal models for the soft emission.
 
An independent estimate of \NH(h) and \NH(w) arises from comparing the 
PSPC and {\it ASCA} fluxes.   The mean fluxes for the joint model, in 
units erg s$^{-1}$ cm$^{-2}$, are
f(c) = 2.88 $\times 10^{-12}$,
f(w) = 1.03 $\times 10^{-12}$, and
f(h) = 1.13 $\times 10^{-12}$.
We can compare these with the PSPC fluxes of
f(c) = $\sim2.0 \times 10^{-12}$,
f(w) = 1.72 $\times 10^{-12}$, and
f(h) = 1.91 $\times 10^{-12}$ erg s$^{-1}$ cm$^{-2}$.
The column density that produces the best agreement between the fluxes 
of the hard and warm components is 4 to 5 $\times 10^{21}$ cm$^{-2}$, 
which is reasonable for a highly-inclined spiral galaxy disk.

\subsection{Summary}

Our conclusions for NGC\,253 are:

(1) There is a large amount of intrinsic absorption in the disk of 
the galaxy (\NH$\sim 1-6\times10^{21}$ cm$^{-2}$).  This absorption 
complicates spectral fits. 

(2) The abundance measurements from {\it ASCA} and PSPC spectra are 
highly uncertain but are statistically consistent with solar values.
Also, the $\alpha$-burning elements are not required to be unusually 
overabundant relative to Fe.

(3) A three-component model that consists of a power law and two MEKAL 
plasma components with solar or near solar abundances provides a fit 
as good or better than a two-component model consisting of a power law 
and a MEKAL plasma having unusual abundances. To maintain consistency 
with spatially-resolved emission, the first alternative is allowed by 
the {\it Rosat} data; the second is not. 

(4) Extrapolating the PSPC results into the {\it ASCA} 
band allows us to infer which spatial components dominate the 
{\it ASCA} spectrum.  The compact sources, 
including the core, provide $30-50\%$ of the 
total $2-10$ keV emission seen with {\it ASCA} and so must make up a 
significant fraction of the $3-10$ keV ``extent'' seen with {\it 
BeppoSAX} (C99) when the telescope point-spread function is accounted 
for.

Comparing our results with those by C99 it is important 
to realize that our three-component model is a composition based
on emission models describing X-ray emitting objects known to exist 
in virtually all spiral galaxies.  While C99 invoke 
the presence of an additional, extremely hot thermal component, they 
neglect all potential continuum emission in the $2-10$ keV band 
from XRBs.  However, there is no reason to believe that NGC\,253,
with its massive starburst, should {\it not} host any X-ray
emitting massive binaries.  On the other hand, XRBs {\it can}
account for the observed thermal Fe K line emission.

\section{Spectroscopic Results for M\,82}
\label{par:m82spec}

\subsection{Modeling the PSPC Spectra}
\label{par:pspc82spec}

\begin{enumerate}

\item The Diffuse Halo Gas:

Power law or single-temperature thermal plasma models are statistically 
ruled out for the halo emission in M\,82 (models P and M, 
Table~\ref{tab:m82_pspcfits}).  Assuming solar abundances, the halo 
emission is best described with a two-temperature model having k$T_1 = 
0.31^{+0.04}_{-0.05}$ keV and k$T_2 = 4.70^{unb}_{-2.82}$ keV, resulting 
in $\chi^2$ = 14.3 for 25 \dof\ (model MM, Table~\ref{tab:m82_pspcfits}). 
As in the case of NGC\,253, we cannot rule out a power law as the harder 
of the two components instead of a thermal plasma, although the thermal 
description is somewhat preferred from the {\it ASCA} result 
(\S~\ref{par:asca82spec}).

\item The Core Spectrum:

Two model components are required to fit the core spectrum.  If these
are a power law and a MEKAL plasma (model PM1), then the hard component 
prefers to be much flatter than measured with {\it ASCA} (similar to 
NGC\,253). This could result from significant unresolved line emission 
above 1 keV (Si) or from a heavily absorbed component.  When the hard 
component is constrained to have the same photon index as the {\it ASCA} 
hard component (model PM2, identical to the technique we used for 
NGC\,253), the inferred thermal temperature in the core approaches 
that in the halo.  In this case, the hard component requires absorption 
in excess of the Galactic value.

\item The Compact Sources:

Most of the point-like sources near M\,82 do not appear to be associated 
with the galaxy since they fall outside the plane of the galaxy.  However, 
the very soft sources to the north have properties similar to the hot 
halo gas in M\,82 and they may, in fact, be associated with the outflow 
(Lehnert, Heckman and Weaver 1999). See paper~1 for further discussion of 
the point sources. 

\end{enumerate}

\subsection{Modeling the {\it ASCA} Spectrum}
\label{par:asca82spec}

{\it ASCA} results for M\,82 have been published by P97, T97, and ML96.
All are of comparable statistical quality, but we propose that the 
spatial information from the PSPC can be used to tell which are 
physically meaningful.

Similar to NGC\,253, we performed a detailed investigation of the {\it 
ASCA} spectrum using a two-component ``baseline'' model to determine 
$\chi^2$(min) and comparing this with the three-component model.  Since 
the statistics are good enough, we also use both Raymond-Smith and 
MEKAL plasma models to see how well the data can distinguish 
between the two. For a model that consists of a thermal plasma with 
solar abundances and a power law, all fits are poor with $\chi^2_\nu$
values ranging from 1.44 to 1.48 (Table~\ref{tab:m82asca}, models 
PR$_{\rm s}$ and PM$_{\rm s}$).  We next allow the abundances of 
individual elements to vary, although keeping N and Na at their solar 
values because of their large uncertainty. In this case, the fits 
improve dramatically (models PR$_{\rm v}$, PM$_{\rm v}$ and BM$_{\rm v}$)
and we derive abundances similar to P97.

Like NGC\,253, the PSPC data clearly indicate an additional low-temperature
plasma and excess absorption (Strickland et al.\ 1997). However, 
three-component models with solar abundances 
(models PRR$_{\rm s}$, PMM$_{\rm s}$) do not fit the {\it ASCA} 
data as well as two-component models with variable abundances.
In fact, when the abundances are 
allowed to vary, a soft component is not required in the {\it ASCA} 
bandpass at all. 

We also detect Fe K emission in M\,82, which could have 
a thermal origin or could come 
from X-ray binaries.  Adding a narrow Gaussian 
to the {\it ASCA} model yields $\Delta\chi^2 = 11$ and a line at 
$6.65\pm0.10$ keV with an equivalent width of $77\pm43$ eV 
and normalization of  
$1.63\pm0.90\times10^{-5}$ photons cm$^{-2}$ s$^{-1}$.   

\subsection{Joint Spectral Modeling}

For joint {\it ASCA} and PSPC fits of M\,82, we are forced to ignore 
the {\it ASCA} data below 0.7 keV to avoid a slight miscalibration 
between the detectors. Currently, there is a calibration effect in 
the SIS at energies less than 1 keV that causes an apparent decrease 
in the quantum efficiency and mimics excess absorption on the order 
of a few $\times10^{20}$ cm$^{-2}$.  The magnitude of the problem 
depends on the SIS CCD mode and the time of the observation since 
launch (Weaver and Gelbord 1999; in prep.).  Early in the mission, 
this effect is only detectable for fairly bright sources and so is 
not a problem for our joint analysis of {\it ASCA} and PSPC 
spectra for NGC\,253 (\S~\ref{par:joint253spec}).

In the case of M\,82 our best fits require the addition of a second 
low-temperature plasma component (models PMM$_{\rm s}$ and PMM$_{\rm v}$, 
Table~\ref{tab:m82jointflux}), regardless of whether or not the abundances 
are allowed to vary.  For variable abundances, O and Mg are uncertain by 
a factor of $\sim$5 and the ratio of O to Fe ranges from 0.92 to 3.6. 
Non-solar abundances are required for some elements, but we find that 
the abundances are not necessarily unusual or overly large when excess 
absorption is allowed.  This result differs from T97 and P97, 
who instead derive 
or assume a low value for \NH(w). Our analysis clearly suggests that the 
relative abundances are model dependent and difficult to determine.

\begin{figure}
\epsfig{file=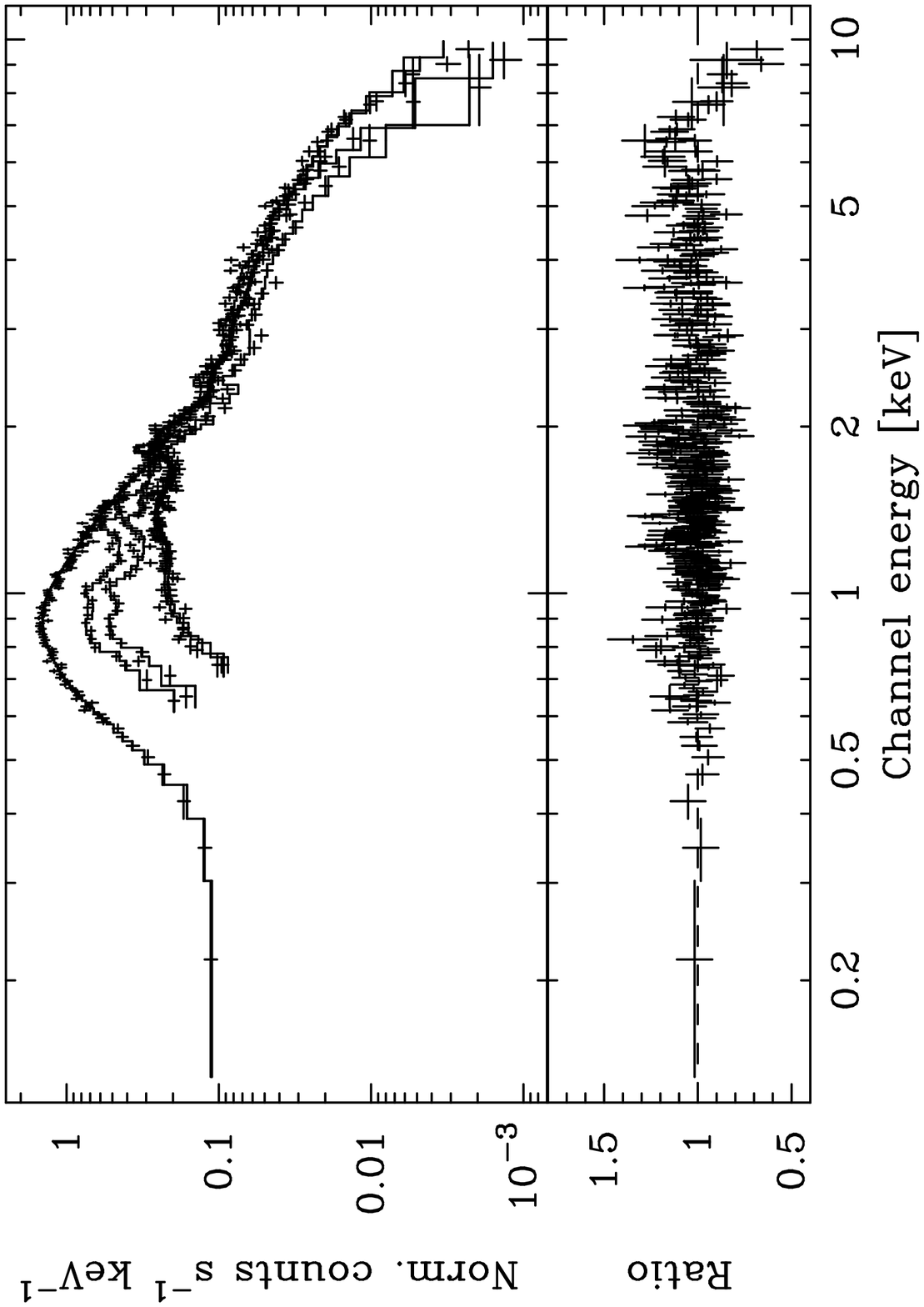, width=6.2cm}
\hspace*{1cm}
\epsfig{file=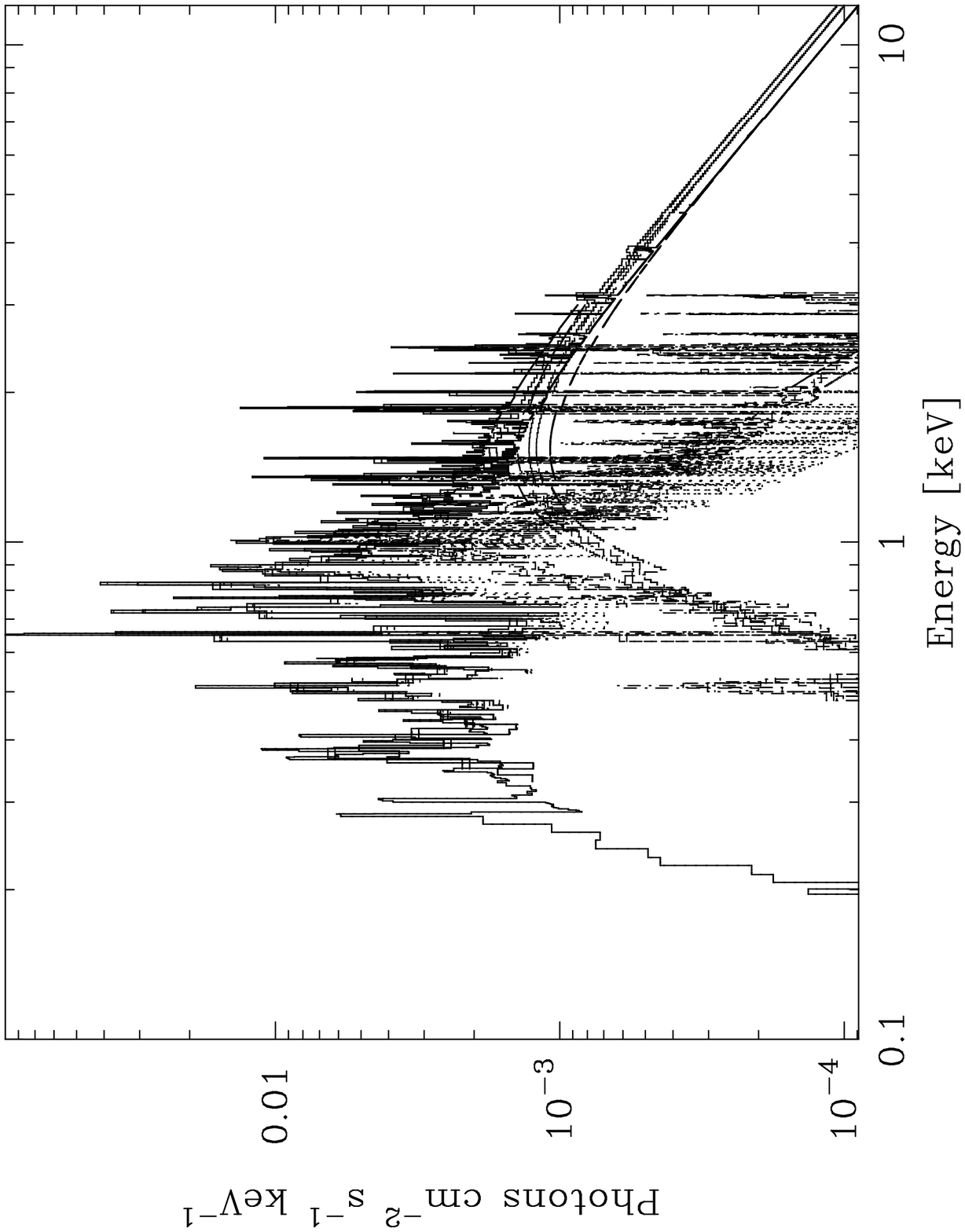, width=6.2cm}
\caption[ ]{(left) {\it ASCA} and {\it Rosat} data for M\,82 and 
the best-fitting two-temperature MEKAL plus power-law model folded through 
the instrumental response (see Table~\ref{tab:m82joint},
model PMM$_{\rm v}$). The 
ratio of the data to this model is shown in the bottom panel.
A Gaussian to model the Fe K emission at 
$\sim6.5$ keV is not included in the model; the line is visible in the 
ratio plot. 
\label{fig:joint82spec}
}
\end{figure}

The most appropriate self-consistent description for the PSPC and {\it 
ASCA} data is a three component model that consists of cool thermal 
emission, warm thermal emission and a hard component.  Model 
PMM$_{\rm v}$ is shown in Figure~\ref{fig:joint82spec}.

\section{Summary \& Implications}

\subsection{Comparison to Previous Work}

We have arrived at somewhat different conclusions than have several
other recent analyses of the X-ray spectra of starbursts. As we have 
argued above, this is not surprising. Starburst galaxies are complex 
X-ray sources, with emission resulting from diffuse, hot gas inside 
and outside the galaxy (associated with starburst-driven winds), X-ray
binaries (XRBs) and supernovae  
in the disk of the galaxy, and absorption from 
the galaxy. The soft X-ray obscuration, in particular when combined with 
limited spatial information, causes ambiguity in unraveling the detailed 
physics of the hot gas (e.g., abundances), because a multi-temperature 
distribution of soft X-ray thermal spectra can be modified by underlying 
absorption and can imitate the signature of gas with unusual abundance 
ratios. 

From {\it ASCA} and {\it Rosat} PSPC archival data for NGC\,253 and 
M\,82, we find ambiguities in the spectral model fitting such that 
multiple components are confused when there is no adequate spatial 
information.  Our most plausible model that is consistent with the 
PSPC images and spectra consists of three components, (1) a 
low-temperature thermal plasma, (2) a warm thermal plasma, and (3) 
a hard power-law or thermal component. We identify the spectral 
components with the likely emitting sources in Table~\ref{tab:spectra}. 
The line-of-sight absorption is approximately a few times 10$^{21}$ 
cm$^{-2}$ in the disk of the galaxy, but both this and the plasma 
abundances are model-dependent, with most constraints being inferred 
from the imaging data.  Our self-consistent examination illustrates 
the criticality of the X-ray images when searching for a physically 
reasonable description of the data.  The absolute abundances are 
impossible to determine without the improved spatial/spectral 
resolution of {\it Chandra} and {\it XMM}.

Our results differ from those by other authors.  T97 examined a 
three component model, but fixed the absorption of the thermal
components to be significantly lower than measured in the
disk (VP98).  They also choose to investigate abundances
and non-equilibrium effects rather than absorption effects, which
we claim have a dramatic impact on the soft X-ray spectrum.
While none of the spectral models used by us, P97, T97, or 
others are unique interpretations of the data, it is worth 
emphasizing that {\it there is no overwhelming evidence at this
time to support highly unusual abundances or non-equilibrium 
effects for M\,82 or NGC\,253}. 

Joint {\it ASCA} + PSPC fits prefer larger absorbing column densities 
than the PSPC alone, except perhaps with a bremsstrahlung modeling of
the hard component.  This appears to be an effect of the different 
ways in which the fits are weighted, with the PSPC fits weighted 
toward the soft emission that is not heavily absorbed and the joint 
fits weighted more toward the warm emission that is absorbed.  We 
expect the joint fits to be more sensitive to intrinsic, underlying 
absorption of the order $10^{21}$ cm$^{-2}$ than the PSPC, because 
such an absorbed component is hard to isolate in the PSPC data.   

The {\it ASCA} data are barely of sufficient quality to comment on 
the spectral origin of the hard X-ray component, i.e., thermal or 
non-thermal.   We detect weak Fe K emission in M\,82, which 
could arise from thermal processes or from XRBs.
There are disagreements about a thermal interpretation for the 
hard X-rays in NGC\,253 because although
{\it BeppoSAX} has now detected the expected Fe K line in NGC\,253, 
the emission is weak (P97 and Ohashi et al.\ 1990).  One explanation 
for the weak Fe lines is that the hard emission contains a significant 
contribution from point sources - XRBs or SNe. Our 
spectral/spatial analysis of 
NGC\,253 suggests that compact sources must at least partly contribute 
to the hard emission.

\subsection{The Origin of the Thermal X-ray Emission in Starbursts}

It has long been known that starbursts are unusually intense X-ray 
sources (cf., Fabbiano 1989 and references therein), and likewise, 
the existence of X-ray-emitting superwinds is now well-established. 
However, the process(es) by which they produce X-ray emission is (are) 
not clear (see, for example Strickland 1998). The chemical abundances 
in the X-ray-emitting gas can provide strong constraints on these 
processes.

If the outflowing wind consists purely of the thermalized ejecta from 
supernovae and stellar winds, the wind fluid would have extremely high 
metallicity ($\sim$ 10 $Z_{\odot}$), with the ratio of $\alpha$-burning 
elements to Fe roughly twice solar (e.g. Gibson, Loewenstein, \& 
Mushotzky 1997). In this case, the gas would be very hot ($\sim$ 10$^8$ 
K prior to adiabatic cooling), but so tenuous that it will be an 
insignificant source of X-ray emission (Chevalier \& Clegg 1985; Suchkov 
et al\ 1994). Thus, the observational constraints on the absolute metal
abundances in the X-ray plasma agree with other arguments: we are
not observing emission from undiluted type II supernova ejecta.

Various processes have been proposed to boost the
X-ray luminosity produced by a superwind. Substantial quantities of
material could be mixed into the wind in or near the starburst as the
stellar ejecta interact with the starburst ISM (the wind could be
centrally ``mass-loaded'' - Suchkov et al\ 1996; Hartquist et al\ 1997).
In this case the abundances will be intermediate between the extreme
values for the supernova ejecta and the more normal values for the gas
in/near the starburst. For example, the hot thermal component we
observe in the cores of NGC\,253 and M\,82 has a temperature of roughly
10$^7$ K, which could imply a roughly 10:1 dilution of the hot
(10$^8$ K) supernova ejecta with ambient gas. In this case, if the
ambient gas had solar abundances, the X-ray gas would have abundances
of $\alpha$-elements like O, Ne, Si, Mg that are roughly twice solar
and only slightly super-solar $\alpha$/Fe ratios. Such abundances are
consistent with our three-component fits to the joint X-ray spectra
of M\,82 and NGC\,253.

Alternatively, very little central mass-loading might occur, and the
hot, tenuous (X-ray faint) wind fluid could propagate into the ISM 
and galactic halo where it could shock heat the halo material to 
10$^6$ to 10$^7$ K, causing it to emit soft X-rays (e.g., Suchkov 
et al\ 1994). In this case we would expect the abundances in the 
X-ray-emitting gas to reflect the pre-existing abundances in the 
ambient gas. In the disk of the ``host'' galaxy or within the starburst
itself, this model would predict roughly solar abundances (consistent
with our three component spectral models). In the halo, substantially
subsolar metal abundances are possible. The High Velocity Clouds in the
halo of the Milky Way generally have metal abundances less than 10\%
solar (Wakker \& van Woerden 1997). The Lyman-Limit quasar absorption-line
systems are generally interpreted as arising in the halos of normal
galaxies (Steidel 1993), and these have abundances ranging from $<$ 0.01 
to 0.4 solar (Prochaska 1999; Prochaska \& Burles 1999). Thus, {\it if} 
the very low metallicities implied by the simple two-component fits are 
correct, it would imply that the observed X-ray emission is arising from
shock-heated halo gas. This may be tenable for the diffuse halo emission
in NGC\,253 and M\,82, but not for the emission in the starburst core.

It is also worth emphasizing that we are now starting to uncover
a complexity in the ISM of starbursts that is similar to that
in the Milky Way. This increasing complexity (e.g., two or three
thermal components in the hot X-ray-emitting gas alone) supports
the suggestions by Norman \& Ferrara (1996) that a more general
description of the ISM (a phase-continuum) should be used instead
of simply adding more and more individual phases. Strickland (1998)
has advocated this approach in fitting the X-ray spectra of starbursts.

\subsection{Starbursts \& the Enrichment of the Intergalactic Medium}

The results in this paper imply that -- contrary to earlier analyses 
-- the hot X-ray-emitting superwind gas can have normal (roughly 
solar) abundances, and therefore may be carrying a significant amount
of metals out of the starburst galaxy and into the intergalactic
medium (IGM). This corroborates the suggestion by Heckman, Armus, \&
Miley (1990; hereafter HAM90) that starburst-driven (i.e., mostly 
dominated by type II supernovae -- hereafter SN II) outflows are one 
of the potential processes for enriching the IGM at the present epoch. 
The importance of this effect in comparison with others is currently a 
matter of debate. While the observed values of the $\alpha$/Fe ratios 
in the intra-cluster medium (ICM) in galaxy clusters leave no doubt 
that most of the metals (by mass) were contributed by SN II (Gibson, 
Loewenstein, \& Mushotzky 1997; Renzini 1997), the roles of Type I 
supernovae, galactic winds, tidal and ram-pressure stripping, elliptical 
vs. spirals, and dwarfs vs. massive galaxies are still matters for 
debate.

For example, Gnedin (1998) argues that tidal-stripping of the ISM is 
the dominant mechanism by which the IGM was ``polluted'', and that
the role of SN II-dominated outflows is negligible on cosmological 
timescales. On the other hand, others have argued that bimodel 
star-formation in elliptical galaxies, i.e., high-mass star-formation 
with an SN II-dominated ISM in the early phase that induces a massive 
wind and dies after about 0.04 Gyr (Arnaud et al. 1992; Elbaz et al\ 
1995) might have caused significant enrichment of the IGM from 
ellipticals then, with no observational trace of further significant 
star-formation at the present epoch (see Zepf \& Silk 1996).

Wiebe et al\ (1999), on the other hand, state that disk galaxies
contribute significantly to the enrichment of the IGM, with an
``effective'' loss (per unit luminosity) comparable to that of
ellipticals, which dominate the enrichment of the IGM only because
they outnumber spirals. The efficiency of metal losses of spirals
into the IGM depends on their total mass. Nath \& Chiba (1995)
find that dwarf starbursts can contribute to the metal enrichment
of clusters at high $z$, but they produce only low-metallicity gas.

Kauffman \& Charlot (1998) have proposed that hot winds of merging
galaxies at high $z$, which later form giant ellipticals, are the
sources of metals in clusters. This idea could logically combine
the two apparently contradicting models favoring metal ingestion
into the IGM by either ellipticals or spirals. The question remains
how significant the metal enrichment of the IGM due to galactic
superwinds in spirals is at the present epoch.

The most likely analogs to these systems in the current universe are far
infra-red luminous galaxies (FIRGs), like e.g. NGC\,6240 and Arp\,220
(HAM90). According to these authors, FIRGs could inject of order $10^9\
M_\odot$ and $10^{59}$ ergs per $L_*$ galaxy over a Hubble time, with no
evolution in the superwind rate over time.  While these systems are
very scarce at low redshifts, sub-millimeter observations show that
dusty ultraluminous galaxies are far more common at high-redshift,
and may rival normal star-forming galaxies in terms of metal
production over the age of the universe (e.g., Lilly 1999). Moreover,
galaxies like M\,82 and NGC\,253 exist in far greater numbers than
ultraluminous mergers in the current universe, and could dominate
the present-day enrichment of the IGM in groups and clusters.

Material torn out of interacting galaxies can be observed in the form of
tidal tails and similar structures (Toomre \& Toomre 1972). The existence
of such structures is a tracer of the shallowness of the galaxies'
gravitational potentials, which also makes it easier for them to expel
metal-enriched gas via supernova-driven superwinds (see, e.g., NGC\,4631;
Weliachew et al\ 1978, Combes 1978, paper~1). In this context it is
important to understand the abundances in these outflows. We have found
in paper~1 that superwinds, with hot gas up to 10 kpc above the disks of
galaxies, do exist and that outflows are a common phenomenon in nearby
starburst galaxies. The current paper shows that the metallicities of
the outflowing gas can be high enough to allow for significant metal
losses into the IGM, if escape velocity is reached (cf., estimates by
HAM90).

The observational limit is currently that the existing ambiguity in the
minimum $\chi^2$ space of the X-ray spectral fits cannot be removed. At
present we can only show that it is indeed possible to fit the data of
both M\,82 and NGC\,253, two of the IR-brightest galaxies in the nearby
Universe, in accordance with chemical evolutionary models and that this
implies possible significant metal losses into the IGM. However, at the
same time spectral fits to the data with extremely low metallicities
are still allowed. As the next step, data of better quality is needed,
first for these nearby galaxies and then for a larger sample of galaxies
to see whether one of the two solutions can then be excluded. This will
be possible with the next generation of X-ray telescopes, like {\it Chandra} 
or {\it XMM}.

X-ray studies of distant (high-$z$) starbursts can be used to estimate
the IGM enrichment by starburst-driven superwinds as a function of
epoch. However, this will also only be possible based on data from one
of the future X-ray missions. A good diagnostic tool is the [O/Fe]
abundance ratio, which should be higher than Solar for SN II-dominated
winds, while for an SN I-dominated ISM it should be close to Solar
(e.g., Matteucci \& Vettolani 1988).
Future high throughput X-ray observations should
put much tighter constraints on the possible metallicities of the hot
gas in star-forming galaxies and thereby help to shed light on the
potential metal enrichment of the IGM as a consequence of massive 
star-formation.

\subsection{Current Limitations \& Future Prospects}

Combining the data from {\it Rosat} PSPC and {\it ASCA} is a step forward 
in our attempt to describe the complex X-ray emission properties of
galaxies like NGC\,253 and M\,82 compared to studies based on data
from only one instrument. Not only is there an ambiguity in the
choice of the ``best'' model for the combined spectra, but an
interpretation of data from only one satellite 
(in the pre-{\it Chandra} and pre-{\it XMM} era) can be even more
misleading because different bandpasses are sensitive to different 
emission components, and neither {\it Rosat} nor {\it ASCA} can detect 
all of them. 

X-ray spectral fits now turn into multi-component multi-d.o.f.\
approximations including internal (X-ray imaging) and external
(observations in other wavebands) consistency checks in the
choice of model components, as performed here. At present, the
choice of models to be fitted to the data depends on
assumptions or information from sources other than the spectra
themselves.

However, there are still a few shortcomings, which call for the
improved capabilities of the next generation of X-ray satellites:

\begin{itemize}

\item
While {\it Rosat} has good spatial resolution but limited spectral 
resolution, {\it ASCA} has a poor spatial resolution but high spectral 
resolution. This leads to possible ambiguities and inconsistencies in
combining the data. The superb  spatial resolution of {\it Chandra},
coupled with an energy resolution similar to {\it ASCA} should allow the
spectral properties to be far better determined, since it will be possible 
to spatially segregate the different emitting components.

\item
The effective areas of the past generations of X-ray telescopes were not 
high enough to provide adequate signal-to-noise for the individual X-ray 
sources/components in starbursts (especially given that M\,82 and 
NGC\,253 are at least an order-of-magnitude brighter X-ray sources 
than any other starburst galaxy). {\it XMM}'s combination of good 
angular resolution, large effective area, and good energy resolution 
will be a major improvement.

\item
For gas in the typical temperature range seen in starbursts, even
the energy resolution of the CCD's on {\it XMM} and {\it Chandra} 
are inadequate to resolve the Fe-L complex (whose strength and mean
energy largely drive the fits for temperature and metal abundance
in the existing spectra). The $\sim$ 10 eV energy resolution of the 
XRS instrument on {\it Astro-E} should allow many individual X-ray 
lines to be resolved. For sources as bright as  M\,82 and NGC\,253,
it may be possible to measure the temperature range and chemical 
abundances via a multitude of independent emission-line ratios (as 
is done in solar/stellar coronae) rather than by fitting idealized 
models to lower-resolution, complex spectra.

\item
Both {\it Chandra} and {\it XMM} cover a very wide energy range
that includes the bandpasses of {\it Rosat + ASCA}, thus removing the 
uncertainties connected with the combination of two datasets 
with disparate properties.

\end{itemize}

{\it Acknowledgements}
We thank the referee, Joel Bregman, for helpful comments.
This research was supported in part by NASA grants
NAG5-6400 and NAG5-6917.

\onecolumn

\tiny{
\begin{table}
\begin{center}
\caption [] {PSPC Spatially-Resolved Spectral Results for NGC\,253}
\label{tab:n253_pspcfits}
\begin{tabular}{lcccccccccc}
\hline
\hline
Region$^a$&Model$^c$&\NH$^d$&kT$^e$&$A_{kT}^f$ & Z$^g$
   &$\Gamma^h$&$A_{\Gamma}^i$&$\chi^2$/$\nu$ $^j$&$f_x$(0.1--2.0keV)$^k$
   &$f_x$(0.1--2.0keV)$^k$ \\
 & & & & & & & & & [Obs.] & [Unabs.] \\
(1) & (2) & (3) & (4) & (5) & (6) & (7) & (8) & (9) & (10) & (11) \\
\hline
 & & & & & & & & & &  \\
Halo &M1&0.90$^{+0.40}$&0.20$^{+0.05}_{-0.07}$&7.0$^{+0.5}_{-0.6}$
  & 1(f)&\nodata&\nodata&96.1/28&1.37$^{+0.09}_{-0.10}$
  &1.85$^{+0.13}_{-0.14}$\\
Halo &M2&0.90$^{+0.40}$&0.36$^{+0.08}_{-0.09}$&49.9$^{+6.3}_{-3.4}$
  &0.01$^{+0.02}_{-0.01}$&\nodata&\nodata&20.1/27&1.35$^{+0.18}_{-0.10}$
  &2.07$^{+0.28}_{-0.14}$\\
Halo &MM&0.90$^{+0.80}$&0.14$^{+0.03}_{-0.06}$&20.0&0.2(f)&\nodata&\nodata
  &19.6/26&0.75(s)&1.29\\
     & & &0.61$^{+0.39}_{-0.39}$&5.69&0.2(f)& & & &0.49(h)&0.60\\
Soft src &P& 1.66$^{+1.74}_{-0.76}$&\nodata&\nodata&\nodata
  &2.90$^{+0.72}_{-0.49}$&0.23$\pm$0.05&14.1/19&0.12$\pm$0.03&0.31$\pm$0.06\\
Soft src &M&0.90$^{+1.09}$&0.40$^{+0.12}_{-0.17}$&4.2$^{+0.7}_{-0.6}$
  &0.0$^{+0.08}_{-0.0}$&\nodata&\nodata&11.2/18&0.12$\pm$0.02&0.18$\pm$0.02\\
Soft src &MM&0.99$^{+2.50}$&0.14$^{+0.04}_{-0.06}$&1.83&0.2(f)&\nodata
  &\nodata&12.1/18&0.07(s)&0.12\\
         & &  & 0.61(f)&0.47&0.2(f)& & & &0.04(h)&0.05\\
 & & & & & & & & & & \\
All Disk&P&9.00$^{+2.5}_{-1.8}$&\nodata&\nodata&\nodata&3.68$^{+0.42}_{-0.36}$
  &3.7$\pm$0.3&34.6/29&1.10$\pm$0.10&17.0$\pm$1.3\\
All Disk&M&4.85$^{+2.25}_{-1.75}$&0.44$^{+0.14}_{-0.11}$&54.5$\pm$4.1
  &0.003$^{+0.025}_{-0.003}$&\nodata&\nodata&37.5/28&0.97$\pm$0.08&2.46$\pm$0.19\\
All Disk&PM1&2.31$^{+1.39}_{-1.11}$&0.25$\pm$0.05&6.09&0.5(f)&0.77(bf)&1.10
  &19.1/27&0.61(s),0.29(h)&0.92(s),0.33(h)\\
All Disk&PM2& 7.02$^{+10.54}_{-3.96}$ & 0.29$\pm$0.03 & 3.83 & 0.5(f) & 1.9(f) & 1.77 
  & 29.1/28 & 0.48(s),0.37(h) & 0.59(s),0.79(h) \\
N. Disk&P&16.5$^{+6.6}_{-4.4}$&\nodata&\nodata&\nodata&3.68(f)&0.72$\pm$0.20
  &31.9/25&0.13$\pm$0.03&3.3$\pm$0.9\\
N. Disk&M&12.0$^{+7.2}_{-4.7}$&0.44(f)&10.4$^{+3.3}_{-2.2}$&0.003(f)
  &\nodata&\nodata&31.1/25&0.12$\pm$0.0&0.47$\pm$0.13\\
S. Disk&P&8.24$^{+0.63}_{-0.52}$&\nodata&\nodata&\nodata&3.68(f)&3.05$\pm$0.22
  &24.1/25&0.96$\pm$0.06&13.8$\pm$1.0\\
S. Disk&M&4.33$^{+0.52}_{-0.43}$&0.44(f)&45.4$\pm$3.2&0.003(f)&
  \nodata&\nodata&20.6/25&0.85$\pm$0.06&2.05$\pm$0.14\\
     & & & & & & & & & & \\
Core&P&5.15$^{+0.67}_{-0.51}$&\nodata&\nodata&\nodata&1.80$^{+0.14}_{-0.12}$
  &8.85$\pm$0.32&108.0/27&2.02$\pm$0.06&3.67$\pm$0.13\\
Core&M&4.29$^{+0.63}_{-0.66}$&1.83$^{+1.17}_{-0.48}$&37.4$^{+1.8}_{-2.0}$
  &0.09$^{+0.17}_{-0.09}$&\nodata&\nodata&92.5/26&1.95$^{+0.09}_{-0.10}$
  &3.01$^{+0.15}_{-0.17}$\\
Core&PM1&2.13$^{+1.25}_{-0.79}$&0.60$^{+0.14}_{-0.15}$&2.90&0.71(bf)
  &0.91$^{+0.41}_{-0.45}$&4.99&27.6/24&0.56(s),1.28(h)&0.65(s),1.46(h)\\
Core&PM2&5.13$^{+0.71}_{-0.57}$&0.80$^{+0.23}_{-0.18}$&1.64&1.0(f)
  &1.9(f)&6.93&70.9/26&0.33(s),1.58(h)&0.42(s),3.06(h)\\
Core&PM3&2.48$^{+1.32}_{-1.23}$&0.51$^{+0.12}_{-0.13}$&13.5
  &0.18$^{+0.12}_{-0.08}$&1.9(f)&10.6&25.2/24&0.90(s),0.89(h)&1.32(s),4.72(h)\\
      & &44.4$^{+48.5}_{-22.1}$& & & & & & & & \\
     & & & & & & & & & & \\
Hard src&P&14.3$^{+13.7}_{-5.8}$&\nodata&\nodata&\nodata
  &2.12$^{+0.80}_{-0.42}$&6.46$^{+0.52}_{-0.46}$&33.1/28&1.03$^{+0.08}_{-0.07}$
  & 3.42$^{+0.28}_{-0.24}$\\
Hard src&B&10.5$^{+5.6}_{-3.7}$&1.86$^{+1.94}_{-0.46}$&8.67$^{+0.67}_{-0.58}$
  &\nodata&\nodata&\nodata&34.5/28&1.02$^{+0.08}_{-0.07}$
  &2.07$^{+0.16}_{-0.14}$\\
Total&P&2.47$^{+0.43}_{-0.37}$&\nodata&\nodata&\nodata&2.23$^{+0.13}_{-0.10}$
  &20.0$\pm$0.06&90.0/28&6.08$\pm$0.12&11.81$\pm$0.30\\
Total&M&1.33$\pm$0.27&1.11$^{+0.20}_{-0.18}$&115.4$\pm$12.0&0.0$^{+0.02}$
  &\nodata&\nodata&64.3/27&5.85$\pm$0.63&7.84$\pm$0.85\\
Total&PM& 0.90$^{+0.4}$&0.34$^{+0.08}_{-0.07}$&22.9$\pm$2.1 
  &0.23(bf)&1.50$^{+0.28}_{-0.27}$
  & 13.3$\pm$0.04 & 21.9/26 & 5.64$\pm$0.60 & 6.90$\pm$0.76 \\
\hline
\end{tabular}
\end{center}
\noindent Notes to Table~\protect\ref{tab:n253_pspcfits}:\\
$^a$Source region (see paper~1).\\
$^b$Background accumulated from field for halo, soft sources, and total;
background accumulated from diffuse halo and galaxy disk emission for 
disk, core, and hard point sources.
For the PSPC spectral analysis we examine both
``field'' and ``local'' backgrounds.
The field background is determined in source-free
areas outside the galaxy that are selected to be larger
than the source regions so as not to be background-noise
limited, while the local background is determined in regions
within the halo and galaxy disk that contain diffuse emission
but that are free of compact sources.  The latter
is important within the central regions
where the spatially-resolved spectra of the galaxy disk
and nucleus can be contaminated by soft, diffuse emission
that is seen in projection. \\
$^c$Models are (P) power law, (M) MEKAL plasma, and (B) thermal bremsstrahlung.
All models include absorption.\\
$^d$Absorbing column density in units of 10$^{20}$ cm$^{-2}$.  Where two values are
listed, the first is for the soft component and the second is for the hard component.\\
$^e$Temperature of thermal component in units of keV.  Where two values are listed
the first is for the soft component and the second is for
the hard component.
For all fits, the absorbing column was constrained to be
greater than the Galactic value of $9\times10^{19}$ cm$^{-2}$.\\
$^f$Normalization of thermal component in units of 10$^{-4}$K, where K is
10$^{-14}$/(4$\pi$ D$^{2}$) ${\int n_{\rm e} n_{\rm H} dV}$ for the MEKAL model and
3.02$\times$10$^{-15}$/(4$\pi$ D$^{2}$) ${\int n_{\rm e} n_{\rm I} dV}$
for the bremsstrahlung model.  D is the luminosity distance to the source in cm and
n$_{\rm e}$, n$_{\rm H}$, and n$_{\rm I}$ are the electron, hydrogen,
and ion densities in cm$^{-3}$.\\
$^g$Abundance in units of $Z_{\odot}$.\\
$^h$Power-law photon index.\\
$^i$Power-law normalization in units of 10$^{-4}$ photons keV$^{-1}$
cm$^{-2}$ s$^{-1}$ at 1 keV.\\
$^j$$\chi^2$ divided by the number of degrees of freedom.\\
$^k$Observed and Unabsorbed $0.1-2.0$ keV fluxes in units of
10$^{-12}$ ergs cm$^{-2}$ s$^{-1}$.  Where
two values are listed, the first is for the soft component and the second is for
the hard component.\\
Except for fluxes and normalizations, which have 90\% confidence errors for
one interesting parameter ($\chi^2$ + 2.71), errors represent 90\% confidence errors
for 2 or 3 interesting parameters ($\chi^2$ + 4.61 and $\chi^2$ + 6.25). \\
``(f)'' denotes a fixed parameter.\\
``(bf)'' denotes a best-fit value that is
fixed to determine errors on the other parameters.\\
\end{table}
}

\clearpage

\tiny{
\begin{table}
\begin{center}
\caption [] {Comparison of {\it ASCA} modeling for NGC\,253$^*$}
\label{tab:n253asca}
\begin{tabular}{lcccccccccccc}
\hline
\hline
Model$^a$& norm$^b$ & $N_{\rm H}$(w)$^c$ & kT(w)$^d$ & norm$^b$ &
$Z_{\rm OMSS}^e$ & $N_{\rm H}$(h)$^c$ &
$\Gamma$/kT$^f$ & norm$^b$ & $\chi^2$/dof & f(c)$^g$ & f(w)$^g$ & f(h)$^g$\\
(1) & (2) & (3) & (4) & (5) & (6) & (7) & (8) & (9) & (10) & (11) & (12) & (13) \\
\hline
PMM & 2.01 & 4.0$^{+1.6}_{-1.9}$ & 0.60$\pm$0.06 & 1.55 &
  2.09$^{+1.04}_{-0.70}$ & 8.3$^{+3.6}_{-3.2}$ & 2.00$^{+0.18}_{-0.18}$
  & 2.04 & 515.4/456 & 1.08 & 1.29 & 1.00 \\
PM & \nodata & 1.5$^{+0.9}_{-1.0}$ & 0.64$\pm$0.07 & 10.7 &
  0.21$\pm$0.08 & 11.7$^{+6.0}_{-4.5}$ & 2.07$^{+0.22}_{-0.20}$ &
  2.28 & 525.9/457 & \nodata & 2.23 & 0.71 \\
BMM & 1.86 & 4.6$^{+1.6}_{-1.5}$ & 0.61$\pm0.06$ &
  1.58 & 2.27$^{+1.24}_{-0.79}$ & 5.0$^{+3.1}_{-2.3}$ &
  7.01$^{+2.74}_{-1.67}$ & 1.59 & 526.0/456 & 1.05 & 1.23 & 1.07 \\
BM & \nodata & 1.3$^{+0.9}_{-0.9}$ & 0.65$\pm0.08$ &
  9.47 & 0.22$^{+0.18}_{-0.07}$ & 7.1$^{+4.0}_{-3.3}$ &
  6.59$^{+2.36}_{-1.69}$ & 1.66 & 540.0/457 & \nodata & 2.19 & 0.78 \\
\hline
\end{tabular}
\end{center}
\noindent Notes for Table~\protect\ref{tab:n253asca}: \\
$^a$Models are M: Mewe-Kaastra plasma (MEKAL), P: power law, B:
bremsstrahlung.  We
assume $Z_{\rm NF}$=$Z_{\rm CN}$=1.0 for 3c and $Z_{\rm NF}$=$Z_{\rm CN}$=0.05
for two-component models, as found by P97.  For the cool component
kT = 0.2 keV and the abundance is 0.2 times solar
(see joint fits).  The Galactic column density is included for all fits.\\
$^b$Normalizations are given in units of 10$^{-3}$. \\
$^c$The units on $N_{\rm H}$ are 10$^{21}$ cm$^{-2}$. \\
$^d$Temperature of the warm component in units of keV.\\
$^e$Abundance of O,Mg,Si, and S in solar units.\\
$^f$Photon index (temperature in keV) of the hard (hot) component.\\
$^g$Fluxes are 0.1-2.0 keV fluxes in units of 10$^{-12}$ ergs cm$^{-2}$ s$^{-1}$.\\
$^*$The spectral extraction regions used here are the same as those of P97. \\
\end{table}
}

\tiny{
\begin{table}
\begin{center}
\caption [] {NGC\,253 {\it ASCA} and PSPC Parameters from Joint Fits}
\label{tab:n253joint}
\begin{tabular}{lcccccccccc}
\hline
\hline
& \multicolumn{2}{c}{Soft} & \multicolumn{5}{c}{Medium} & \multicolumn{2}{c}{Hard} &  \\
Model$^a$ & $kT_1^b$ & $Z^c$&$N_{\rm H}^d$(w)&$kT_2^e$&$Z_{\rm HCN}$&
$Z_{\rm OMSS}$&$Z_{\rm NF}$&$N_{\rm H}^f$(h)&$\Gamma$/kT&$\chi^2/\nu$ \\
 (1) & (2) & (3) & (4) & (5) & (6) & (7) & (8) & (9) & (10) & (11) \\
\hline
PMM1    & 0.26$\pm$0.04 & 0.18$^{+0.11}_{-0.05}$ & 6.38$^{+2.94}_{-3.65}$  
  & 0.70$\pm$0.10 & 0.5(f) & 0.8$^{+1.9}_{-0.5}$ & 0.4$^{+1.6}_{-0.3}$  
  & 8.51$^{+3.54}_{-3.05}$ & 1.96$^{+0.18}_{-0.20}$ & 538.5/499\\
PMM2    & 0.26$\pm$0.04 & 0.17$^{+0.11}_{-0.05}$ & 6.77$^{+3.03}_{-3.97}$
  & 0.72$\pm0.09$ & 0.5(f) & 0.87$^{+1.23}_{-0.47}$ &
  0.5(f) & 7.91$^{+3.99}_{-3.26}$ & 1.95$^{+0.17}_{-0.19}$ & 540.5/500 \\
BMM     & 0.27$\pm$0.04 & 0.17$^{+0.10}_{-0.05}$ & 7.67$^{+3.23}_{-4.62}$
  & 0.75$\pm$0.09 & 0.5(f) & 0.96$^{+2.34}_{-0.58}$ &
  0.5(f)& 4.60$^{+4.30}_{-2.63}$ & 7.8$^{+4.7}_{-2.0}$ & 544.2/500 \\
PMM3    & 0.27$\pm$0.03 & 0.17$^{+0.09}_{-0.05}$ & 7.55$^{+2.85}_{-3.25}$
  & 0.74$^{+0.16}_{-0.14}$ & 0.5(f) & 0.79$^{+1.42}_{-0.45}$ & 0.5(f)&
  =NH(m) & 1.92$^{+0.14}_{-0.16}$ & 540.8/501 \\
PMM4    & 0.27$\pm$0.03 & 0.17$^{+0.09}_{-0.05}$ & 7.28$^{+1.40}_{-2.78}$
  & 0.73$^{+0.11}_{-0.14}$ & 1.0(f) & 1.67$^{+0.67}_{-0.75}$ & 1.0(f)&
  =NH(m) & 1.94$^{+0.10}_{-0.11}$ & 540.6/501\\
(PMM)pc & 0.34$\pm$0.07 & $=Z(hard)$ & 3.76$^{+4.74}_{-3.67}$ &
  0.72$^{+0.30}_{-0.13}$ & 1.0(f) & 3.35$^{+3.91}_{-2.17}$ & 1.0(f)
  & $^*$12.9$^{+4.3}_{-2.9}$ & 2.12$^{+0.18}_{-0.17}$ & 540.5/500\\
\hline
\end{tabular}
\end{center}
\noindent Notes to Table~\protect\ref{tab:n253joint}:\\
$^a$Model components are P:power law, 
B: bremsstrahlung, M: MEKAL plasma, pc: partial covering. \\
$^b$Temperature of cool component in units of keV.\\
$^c$Abundance of cool component in units of \zsol.\\
$^d$Column density of warm component in units of 10$^{21}$ cm$^{-2}$.\\
$^e$Temperature of warm component in units of keV.\\
$^f$Column density of hard component in units of 10$^{21}$ cm$^{-2}$.\\
Errors for NH(w), NH(h), $\Gamma$, and z(OMSS) are 90\% confidence for 4 interesting
parameters ($\Delta\chi^2$=7.1).  Errors for kT(m) are $\Delta\chi^2$=4.6 for cases
where both columns are free.\\
The Galactic column of 0.9$\times$10$^{20}$ included in all fits.\\
$^*$For the partial covering model, the covering factor is $f_c = 0.77^{+0.08}_{-0.13}$\\
\end{table}
}

\small{
\begin{table}
\begin{center}
\caption [] {NGC\,253 Normalizations and fluxes from Joint Fits}
\label{tab:n253jointflux}
\begin{tabular}{lcccccc}
\hline
\hline
Model & A(c)$^a$ & A(w)$^a$ & A(h)$^a$ & f(c)$^b$ & f(w)$^b$ & f(h)$^b$ \\
(1) & (2) & (3) & (4) & (5) & (6) & (7) \\
\hline
PMM1  & 5.33$^{+0.85}_{-1.04}$ & 4.27$^{+1.62}_{-1.24}$ & 2.05$^{+0.38}_{-0.29}$ & 
   2.93 & 1.04 & 1.04 \\
PMM2  & 5.48$^{+0.80}_{-0.97}$ & 3.64$^{+1.55}_{-1.19}$ & 2.00$^{+0.31}_{-0.26}$ &
   2.96 & 0.95 & 1.09 \\
BMM  & 5.61$^{+0.75}_{-0.89}$ & 3.72$^{+0.85}_{-0.67}$ & 1.62$^{+0.14}_{-0.12}$ &
   2.99 & 0.87 & 1.15 \\
PMM3  & 5.58$^{+0.21}_{-0.19}$ & 4.03$^{+1.32}_{-0.74}$ & 1.90$^{+0.20}_{-0.19}$ &
   3.00 & 0.90 & 1.09 \\
PMM4  & 5.54$^{+0.21}_{-0.20}$ & 1.94$^{+0.17}_{-0.16}$ & 1.98$\pm$0.15 & 2.98 &
   0.86 & 1.16 \\
(PMM)pc & 0.34$\pm$0.10 & 0.78$^{+0.13}_{-0.14}$ & 2.69$\pm$0.33 & 1.60 & 0.88 & 3.17\\
\hline
\end{tabular}
\end{center}
\noindent Notes to Table~\protect\ref{tab:n253jointflux}:\\
$^a$Normalizations for the cool, warm, and hard components
are given in units of 10$^{-3}$.\\
$^b$Fluxes are 0.1-2.0 keV fluxes in units of 10$^{-12}$ ergs cm$^{-2}$ s$^{-1}$.\\
\end{table}
}

\tiny{
\begin{table}
\begin{center}
\caption [] {PSPC Spectral Results for M\,82}
\label{tab:m82_pspcfits}
\begin{tabular}{cccccccccccc}
\hline
\hline
(1) & (2) & (3) & (4) & (5) & (6) & (7) & (8) & (9) & (10) & (11) & (12) \\
Region & M,F & \NH & k$T$ & $A_{\rm kT}$ & $Z/Z_{\odot}$ & $\Gamma$ &
  $A_{\Gamma}$ & $\chi^2$/$\nu$ & $f_x$ & $f_x$ & $L_x$ \\
&  & [$10^{20}$ cm$^{-2}$] & [keV] & [10$^{-3}$] & & & [10$^{-3}$] & & [Obs.]
  & [Unabs.] & [Unabs.] \\
(1) & (2) & (3) & (4) & (5) & (6) & (7) & (8) & (9) & (10) & (11) & (12) \\
\hline
Halo & P & 47.0$^{+17.0}_{-11.0}$ & \nodata & \nodata & \nodata &
 6.32$^{+1.20}_{-0.87}$ & 2.19$^{+0.16}_{-0.17}$ & 77.0/27 &
 1.58$^{+0.11}_{-0.07}$ & \nodata & \nodata \\
Halo & M & 7.5$^{+1.5}_{-1.6}$ & 0.45$\pm$0.06 & 8.10$\pm$0.40 &
 0.03$\pm$0.02 & \nodata & \nodata & 34.9/26 & 1.58$\pm$0.07 &
 4.27$\pm$0.21 & 5.42$\pm$0.26 \\
Halo & MM &4.4$^{+2.0}_{-0.7}$ & 0.31$^{+0.04}_{-0.05}$ &
 0.57$\pm$0.05 & 1.0(f) & \nodata & \nodata & 14.3/25 & 1.50$\pm$0.15 &
 2.26$\pm$0.19 & 2.87$\pm$0.25 \\
& & & 4.70$^{+unb}_{-2.82}$ & 0.74$\pm$0.10 & 1.0(f) & & & & & & \\
Core & P & 33.5$^{+3.5}_{-2.6}$ & \nodata & \nodata & \nodata &
 2.98$^{+0.19}_{-0.18}$ & 12.5$^{+0.2}_{-0.1}$ & 502.1/27 &
 10.8$^{+0.3}_{-0.3}$ & \nodata & \nodata \\
Core & PM1 & 7.6$^{+2.3}_{-1.3}$ & 0.67$^{+0.04}_{-0.05}$ &
 1.92$^{+0.13}_{-0.12}$ & 1.0(f) & 0.39$^{+0.26}_{-0.29}$ &
 2.98$^{+0.27}_{-0.26}$ & 41.2/25 & 11.1$\pm$1.0 &
 15.1$\pm$1.2 & 19.18$\pm$1.52 \\
Core & PM2 & 31.8$^{+12.1}_{-5.6}$ & 0.54$^{+0.10}_{-0.15}$ &
 1.92$^{+1.15}_{-0.45}$ & 1.0(f) & 1.50(f) & 5.37$^{+0.33}_{-0.26}$ &
 46.3/28 & 8.56$^{+1.46}_{-0.91}$ &
 26.5$^{+4.51}_{-2.82}$ & 33.7$^{+5.7}_{-3.6}$ \\
Total& P & 31.4$^{+3.2}_{-3.3}$ &\nodata&\nodata&\nodata&3.25$^{+0.2}_{-0.26}$ &
  13.7$^{+1.4}_{-1.1}$ & 568.3/166 & 12.4$^{+1.3}_{-1.3}$ & & \\
Total& M &16.4$^{+4.7}_{-1.4}$ & 0.97$^{+0.08}_{-0.17}$ & 52.0$^{+16.0}_{-4.0}$ & 
  0.02$\pm0.01$ & \nodata & \nodata &
  540.2/165 & 12.3$^{+3.7}_{-0.5}$ & & \\
Total& PM & 5.7$^{+0.8}_{-0.7}$ & 0.60$^{+0.05}_{-0.03}$ & 2.08$^{+0.13}_{-0.14}$ & 
  1.0(f) & 0.57$^{+0.09}_{-0.08}$ & 3.34$^{+0.35}_{-0.33}$ &
  124.7/164 & 12.7$^{+1.0}_{-1.5}$ & & \\
\hline
\end{tabular}
\end{center}
\noindent Notes to Table~\protect\ref{tab:m82_pspcfits}:\\
Models, normalizations and errors are the same as in
Tables~\ref{tab:n253joint} and \ref{tab:n253jointflux}.\\
Fluxes are 0.1--2.0 keV fluxes in units of [10$^{-12}$ ergs cm$^{-2}$
s$^{-1}$]. Luminosities are 0.1--2.0 keV luminosities in units of
[10$^{39}$ ergs s$^{-1}$].\\
The core spectrum has local background subtracted.\\
\end{table}
}

\tiny{
\begin{table}
\begin{center}
\caption [] {Comparison of {\it ASCA} modeling for M\,82}
\label{tab:m82asca}
\begin{tabular}{lcccccccccccc}
\hline
\hline
Model$^a$& norm$^b$ & $N_{\rm H}$(m)$^c$ & kT(w)$^d$ & norm$^b$ & $Z$
& $N_{\rm H}$(h)$^c$ &
$\Gamma$/kT$^f$ & norm$^b$ & $\chi^2$/dof & f(c)$^g$ & f(w)$^g$ & f(h)$^g$\\
(1) & (2) & (3) & (4) & (5) & (6) & (7) & (8) & (9) & (10) & (11) & (12) & (13) \\
\hline
PR$_{\rm s}$ & \nodata & 7.9$^{+4.1}_{-4.5}$ & 0.77$\pm$0.03 & 1.22$\pm$0.16
     & 1(f) &\nodata & 1.50$\pm$0.04 & 3.58$\pm$0.05
     & 1205.0/813 &\nodata& 8.5$\pm$0.2 & 3.0$\pm$0.4 \\
PM$_{\rm s}$ & \nodata & 2.1$^{+0.6}_{-0.5}$ & 0.60$\pm$0.03 & 2.70$\pm$0.11
     & 1(f)& \nodata & 1.55$\pm$0.04 & 3.91$\pm$0.10
     & 1167.7/813 &\nodata & 4.0$\pm$0.2 & 6.5$\pm$0.3 \\
PR$_{\rm v}$ & \nodata & 1.1$^{+0.5}_{-0.3}$ & 0.75$\pm$0.05 & 29.5$\pm$1.4
     &(a)& 25.5$^{+3.6}_{-7.7}$ & 1.77$\pm$0.08 & 6.19$\pm$0.10
     & 903.2/805 &\nodata & 11.1$\pm$0.6 & 1.0$\pm$0.7 \\
PM$_{\rm v}$ & \nodata & 2.2$^{+0.3}_{-0.4}$ & 0.61$\pm$0.04 & 32.3$\pm$2.2
     &(b)&16.7$^{+3.9}_{-3.3}$ & 1.71$\pm$0.10 & 5.42$\pm$0.09
     & 879.3/804 &\nodata & 9.7$\pm$0.8 & 1.5$\pm$0.1 \\
BM$_{\rm v}$ & \nodata & 2.1$^{+0.3}_{-0.4}$ & 0.61$\pm$0.03 & 29.4$\pm$2.4
     &(c)&11.6$^{+3.0}_{-2.7}$ & 15.4$^{+5.0}_{-3.5}$&5.44$\pm$0.08
     & 875.7/804 &\nodata & 9.4$\pm$0.8 & 1.7$\pm$0.1 \\
$^*$PRR$_{\rm s}$ & 1.31$\pm$0.07 & 7.60$^{+0.8}_{-0.9}$ & 0.77$\pm$0.03 & 12.6$\pm$0.3
     & 1(f)& 8.9$^{+1.8}_{-1.6}$ & 1.47$\pm$0.08 & 3.50$\pm$0.10 & 969.6/811
     & 3.0$\pm$0.2 & 5.5$\pm$0.2 & 2.3$\pm$0.1 \\
$^*$PMM$_{\rm s}$ & 1.53$\pm$0.07 & 8.40$\pm$0.9 & 0.67$\pm$0.04 & 16.7$\pm$0.4
     & 1(f)& 10.6$^{+3.5}_{-3.1}$ & 1.54$\pm$0.09 & 3.96$\pm$0.06 & 952.4/811
     & 3.3$\pm$0.2 & 5.3$\pm$0.2 & 2.2$\pm$0.1 \\
\hline
\end{tabular}
\end{center}
\noindent Notes for Table~\protect\ref{tab:m82asca}:\\
$^a$Models are R: Raymond-Smith plasma, P: power law, B:
bremsstrahlung.\\
*For the cool component,
the temperature is fixed at 0.31 keV and $N_{\rm H} = 3.71 \times 10^{20}$
cm$^{-2}$. \\
(a) abundances are: N=1(f),O=0.0, Ne=0.16, Mg=0.26,
Si=0.35, S=0.60, Ar,Ca,Ni=0.014, Fe=0.034 (Si/Fe = 10.3). \\
(b) abundances are: N=1(f), O=0.19, Ne=0.20, Na=1(f), Mg=0.33,
Al=0.0, Si=0.45, S=0.84, Ar,Ca,Ni=0.0, Fe=0.07 (Si/Fe = 6.4).\\
(c) abundances are: N=1(f), O=0.21, Ne=0.23, Na=1(f), Mg=0.34,
Al=0.0, Si=0.49, S=0.97, Ar,Ca,Ni=0.0, Fe=0.08 (Si/Fe = 6.1).\\
Fluxes are averaged GIS fluxes for M\,82.\\
$^*$\NH(c) = $6\times10^{20}$ cm$^{-2}$ and kT(c) = 0.35 keV.\\
\end{table}
}

\tiny{
\begin{table}
\begin{center}
\caption [] {M\,82 {\it ASCA} and PSPC Parameters from Joint Fits}
\label{tab:m82joint}
\begin{tabular}{lccccccccc}
\hline
\hline
&\multicolumn{3}{c}{Soft} &\multicolumn{3}{c}{Medium} &\multicolumn{2}{c}{Hard}& \\
Model$^a$ & $N_{\rm H}^d$(c)& $kT_1^b$ & $Z^c$&$N_{\rm H}^d$(w)&$kT_2^e$&$Z^c$
  &$N_{\rm H}^f$(h)&$\Gamma$/kT&$\chi^2/\nu$ \\
 ~(1) &  (2) & (3) & (4) & (5)  & (6) & (7) & (8) & (9) & (10) \\
\hline
PR$_{\rm s}$ &3.7$^{+1.5}$ &\nodata&\nodata&\nodata& 0.76$^{+0.02}_{-0.03}$ & 1.0(f)
      & 8.1$^{+1.7}_{-5.3}$ & 1.54$^{+0.03}_{-0.04}$    & 1396.3/975 \\
PR$_{\rm v}$ &4.6$^{+1.1}_{-0.8}$ &\nodata&\nodata&\nodata& 0.59$\pm$0.03 & (a)
      & 1.59$^{+0.76}_{-0.61}$ & 1.44$^{+0.05}_{-0.06}$ & 1095.1/968 \\
PM$_{\rm s}$ &3.8$^{+0.6}_{-0.1}$ &\nodata&\nodata&\nodata& 0.64$\pm$0.02 & 1.0(f)
      & 14.$\pm$3.9 & 1.58$^{+0.03}_{-0.04}$            & 1360.9/975\\
PM$_{\rm v}$ &5.72$^{+1.2}_{-0.9}$ &\nodata&\nodata&\nodata& 0.60$\pm$0.03 & (b)
      & 2.4$^{+0.9}_{-0.8}$ & 1.48$^{+0.05}_{-0.07}$    & 1066.6/968 \\
PMM$_{\rm s}$ &5.3$^{+1.0}_{-0.8}$&0.33$\pm$0.03&1.0(f) & 8.7$^{+0.6}_{-0.7}$
    & 0.68$\pm$0.05
    & 1.0(f)&10.1$^{+3.6}_{-2.4}$&1.53$^{+0.09}_{-0.08}$& 1110.9/972\\
PMM$_{\rm v}$ &6.2$^{+1.0}_{-0.9}$ & 0.48$\pm$0.05 &(d)& 4.5$^{+1.6}_{-1.4}$
    & 0.92$^{+0.13}_{-0.12}$
    & (d) & 6.1$^{+1.7}_{-1.8}$ & 1.47$\pm$0.09         & 1045.5/965\\
\hline
\end{tabular}
\end{center}
\noindent Notes to Table~\protect\ref{tab:m82joint}:\\
$^a$Model components are P: power law, R: Raymond-Smith plasma,
M: MEKAL plasma.  The s and v subscripts refer to solar and
variable abundances, respectively.\\
$^b$Temperature of cool component in units of keV.\\
$^c$Abundance in units of \zsol.\\
$^d$Column density of warm component in units of 10$^{21}$ cm$^{-2}$.
Column density of soft component in units of 10$^{20}$ cm$^{-2}$.\\
$^e$Temperature of warm component in units of keV.\\
$^f$Column density of hard component in units of 10$^{21}$ cm$^{-2}$.\\
Errors for NH(w), NH(h), $\Gamma$, and Z are 90\% confidence for 4 interesting
parameters ($\Delta\chi^2$=7.1).  Errors for kT(w) are $\Delta\chi^2$=4.6 for cases
where both columns are free.\\
The Galactic column is 3.7$\times$10$^{20}$ cm$^{-2}$.\\
(a) abundances are: N=1(f), O=3.17, Ne=4.47, Mg=6.41, Si=13.6,
S=33.8, Ar,Ca,Ni=8.93, Fe=0.88.\\
(b) abundances are: N=1(f), O=4.34, Ne=5.63, Na=1(f), Mg=9.43, Si=15.5,
S=29.5, Ar,Ca,Ni=4.88, Fe=1.98.\\
(c) abundances are: N=1(f), O=1.94, Ne=4.70, Mg=3.97, Si=3.58,
S=5.26, Ar,Ca,Ni=8.98, Fe=2.10.\\
(d) abundances are: N=1(f), O=0.85, Ne=0.01, Na=1(f), Mg=1.78, Si=2.05,
S=2.41, Ar,Ca,Ni=3.96, Fe=0.51.\\
\end{table}
}

\small{
\begin{table}
\begin{center}
\caption [] {M\,82 Normalizations and fluxes from Joint Fits}
\label{tab:m82jointflux}
\begin{tabular}{lcccccc}
\hline
\hline
Model & A(c)$^a$ & A(w)$^a$ & A(h)$^a$ & f(c)$^b$ & f(w)$^b$ & f(h)$^b$ \\
\hline
PR$_{\rm s}$ & \nodata & 1.39$\pm$0.08 & 4.98$\pm$0.05 & \nodata& 3.5 & 8.9 \\
PR$_{\rm v}$ & \nodata & 1.11$\pm$0.05 & 4.24$\pm$0.04 & \nodata& 5.6 & 5.5 \\
PM$_{\rm s}$ & \nodata & 1.75$\pm$0.07 & 5.36$\pm$0.05 & \nodata& 4.1 & 8.2 \\
PM$_{\rm v}$ & \nodata & 1.02$\pm$0.04 & 4.54$\pm$0.04 & \nodata& 6.2 & 5.1 \\
PMM$_{\rm s}$ & 2.35$\pm$0.10 & 21.1$\pm$0.6  & 4.92$\pm$0.07 & 4.8 & 5.9 & 2.9 \\
PMM$_{\rm v}$ & 3.78$\pm$0.15 & 5.54$\pm$0.14 & 4.54$\pm$0.06 & 6.1 & 2.9 & 3.4 \\
\hline
\end{tabular}
\end{center}
\noindent Notes to Table~\protect\ref{tab:m82jointflux}:\\
$^a$Normalizations for the cool, warm, and hard components
are given in units of 10$^{-3}$.\\
$^b$Fluxes are $0.1-2.0$ keV
observed fluxes in units of 10$^{-12}$ ergs cm$^{-2}$ s$^{-1}$.\\
\end{table}
}

\small{
\begin{table}
\begin{center}
\caption [] {Spectral Models for Spatially-Resolved Emission}
\label{tab:spectra}
\begin{tabular}{lccccc}
\hline
\hline
Galaxy  & Halo & Disk & Core & Pt. sources & Notes \\
(1) & (2) & (3) & (4) & (5) & (6) \\
\hline
NGC\,253 & M+M & M+M & P+M & P or M &  absorption in disk\\
M\,82  & M+M or P+M & \nodata & P+M & P or M & absorption in disk and halo \\
\hline
\end{tabular}
\end{center}
\noindent Notes to Table~\protect\ref{tab:spectra}: \\
Models are M: Mekal plasma, P: power law.\\
\end{table}
}


%

\end{document}